\documentclass[11pt]{amsart}
\usepackage[cp1251]{inputenc}
\usepackage{graphicx,color}
\usepackage[left=1in,right=1in,top=1in,bottom=1in]{geometry}
\usepackage{latexsym}
\usepackage{cite}
\usepackage{amsmath,amssymb,amsthm,mathrsfs,amscd}

\allowdisplaybreaks[1]  

\newtheorem{theorem}{Theorem}

\newcounter{mnotecount}[section]
\let\oldmarginpar\marginpar
\renewcommand\marginpar[1]{\-\oldmarginpar[\raggedleft\footnotesize #1]%
{\raggedright\footnotesize #1}}

\begin{document}

%
%
%
%
%
%
%
%
%

\title{A scale-dependent distance functional between past-lightcones in cosmology}

\author[M. Carfora]{Mauro Carfora}

\address[Department of Physics, University of Pavia]{University of Pavia} 
\address[GNFM and INFN]{Italian National Group of Mathematical Physics, and INFN Pavia Section}
\email{mauro.carfora@unipv.it}

\author[F. Familiari]{Francesca Familiari}

\address[Department of Physics, University of Pavia]{University of Pavia} 
\address[GNFM and INFN]{Italian National Group of Mathematical Physics, and INFN Pavia Section} 

\email{francesca.familiari01@universitadipavia.it}





\begin{abstract}
We discuss a rigorous procedure for quantifying the difference between our past lightcone and the past lightcone of the fiducial Friedmann-Lemaitre-Robertson-Walker  spacetime modeling the large scale description of cosmological data in the standard $\Lambda\mathrm{CDM}$ scenario. This result is made possible by exploiting the scale-dependent distance functional between past lightcones recently introduced by us in \cite{CarFam}. We  express this harmonic map type functional  in terms of the physical quantities that characterize the actual measurements along our past lightcone, namely the area distance and the lensing distortion,  also addressing the very delicate problem of  the presence of  lightcone caustics.  This analysis works beautifully and seems to remove several of the difficulties encountered in comparing the actual geometry of our past lightcone with the geometry of the fiducial FLRW lightcone of choice.  We also discuss how, from the point of view of the FLRW geometry,  this distance functional may be interpreted as a  scale-dependent effective field, the pre-homogeneity field,  that may be of relevance in selecting the FLRW model that best fits the observational data.
\end{abstract}

\maketitle

\section{INTRODUCTION}
\label{Introduction}
\noindent
\emph{It is a pleasure to dedicate this paper to Maurizio Gasperini
who has always liked it best on the\\ past light cone even if
the routes are tough,
but in such a rugged landscape that is to be expected}\\
\\
\noindent
The  $\Lambda\mathrm{CDM}$ model and the Friedman-Lemaitre-Robertson-Walker (FLRW) spacetimes  provide a rather accurate physical and geometrical representation of the universe in the present era\footnote{Characterized by the actual temperature of the cosmic microwave background $T_{CMB}\,=\,2.725\,K$ as measured  in the frame centered on us
but stationary with respect to the CMB.} and  over spatial scales ranging from\footnote{
The actual averaging scale marking the statistical onset of isotropy and homogeneity is still much debated. For  the sake of the argument presented in this paper, we adopt the rather conservative estimate of the  scales over which an average
isotropic expansion is seen to emerge, namely $70-120\,h^{-1}Mpc$, and ideally extending to a few times this scale \cite{Wiltshire}.} $\approx  \, 100\,h^{-1}\; \mathrm{Mpc}$ to the visual horizon of  our past light cone \cite{Gott}, \cite{Hogg}, \cite{Scrimgeour}, where $h$ is the dimensionless  parameter  describing  the relative uncertainty of the true value of the present-epoch  Hubble-Lemaitre constant.  Within such observational range, and on scales significantly smaller than the Hubble scale\footnote{At the Hubble scale, the problem of \emph{cosmic variance} may  alter the statistical significance of the data samples we gather.}, we have a testable  ground for  statistical isotropy in the distribution of the dark and visible matter components on  our past light cone. Homogeneity of this distribution is difficult to test directly via astronomical surveys, but a number of  observational results \cite{Maartens} and in particular the kinematic Sunyaev-Zeldovich effect \cite{Zhang}, \cite{EllisF} imply that  fluctuations around spatial homogeneity cannot be too large. Thus, without resorting to an axiomatic use of  the Copernican principle, we have an observational ground for assuming that spatial homogeneity holds, in a statistically averaged sense, over large scales. It must be stressed that it is in a statistical sense and only over large scales that this weak form of the cosmological principle provides observational support for best fitting the description of spacetime geometry in terms of a member of  the FLRW family of solutions of the Einstein equations. In particular, to whatever degree one accepts this FLRW scenario, one has to address the fact that the role of  FLRW spacetime geometry becomes  delicate to interpret when past light cone data are gathered  in our cosmological neighborhood.  As we probe spatial regions  in the  range $\lesssim  \, 100h^{-1}\; \mathrm{Mpc}$, the actual distribution of matter (dark and visible) becomes extremely anisotropic with a high density contrast. In particular,  gravitational clustering gives rise to a complex network of structures, characterized by the presence of  a foam-like web of voids and galaxy filaments often extending well into the  $100h^{-1}\; \mathrm{Mpc}$ range.  At these scales,  the Einstein evolution of  the  FLRW geometry uncouples from the dynamics of the matter sources and survives more as a useful computational assumption (often assisted by Newtonian theory) rather  than as a  \emph{bona fide} perturbative background gravitationally determined by the actual matter distribution. FLRW is thus a very strong assumption and not a correct representation of spacetime geometry at the pre-homogeneity scales, not even in a statistical sense.  If we want or need to go beyond FLRW perturbation theory and enter into a fully relativistic regime,  it is  fair to say that we have little  mathematical control over the actual  spacetime at these pre-homogeneity scales.  In particular,  the transition from  the large-scale FLRW  to the actual inhomogeneous and anisotropic spacetime geometry emergent at these local scales is  poorly understood in a model-independent way, and the idea that 
around $100h^{-1}\; \mathrm{Mpc}$ we have a gradual and smooth transition between these two regimes is somewhat illusive.  To wit,  we may have non-perturbative correction terms due to the coupling between gravitationally bound structures and the emergent spacetime geometry  (\textit{e.g.} structure formation-induced curvature) that can be significant in  cosmological modeling. For instance,  they can back-react, in a top-down causation way \cite{Uzan}, on the  choice of  the large-scale FLRW spacetime that best fits  the  observational data. This complex scenario gives rise to a number of  delicate and to some extent controversial issues that are currently much debated in discussing  the existence of possible tensions between cosmological observations and the standard $\Lambda\mathrm{CDM}$ model and in preparation to the coming era of high-precision cosmology \cite{Buchert}. Some of the very delicate reasons\footnote{We wish to thank one of the referees for pointing this out to us.} motivating this tension is that large-scale isotropy can hold for a much wider class of models, the so-called \emph{effective model} \cite{Heinesen}, \cite{Heinesen2} that need not even be a solution of Einstein's equations. As an illustrative example, one may
consider inhomogeneous spatial sections that can be smoothed into a constant-curvature space, \emph{e.g.} with Ricci flow deformation techniques \cite{Carfora_2, Carfora_1, Carfora0, Carfora1, Carfora2, Carfora3}. While
spatially, such slices can be identified with spatial sections of a FLRW model, their Einstein time-evolution in general does not follow the FLRW class of solutions, a  backreaction is present \cite{Carfora0, Carfora1, Carfora2}. Thus, at least in principle, one may actually deal with an effective model with global backreaction that can be large-scale isotropic and homogeneous, or almost so,
and it is not necessarily perturbatively away from a FLRW model. Thus, restricting a priori the "best-fit" to the class of FLRW models is indeed a strong assumption.
By its very nature, a discussion of this very complex scenario should
be related, as far as possible, to a model--independent \emph{direct observational cosmology approach}, namely to the analysis of data determined on our past lightcone without using any theory of gravity. Since the dark matter and dark energy  components cannot be measured yet via direct observations, it must be stressed that a full  
 model-independent cosmographic approach is not actually possible \cite{Ellis2}. Model hypotheses must be imposed for the dark components, in particular on how they interact with observed matter.  The simplest assumptions made  are that the dark matter component follows the baryonic component,\,namely that: \textit{i)}\,we know the primordial ratio of cold dark matter (CDM) density to baryonic density; \textit{ii)}\, they have the same 4-velocity; \textit{iii)}\, we know their relative concentration in matter clusters. To these, one typically adds the working assumption that the dark energy component is described in the form of a cosmological constant $\Lambda$, the value of which should be known from non--cosmological physics and independently from cosmological observations (for a thorough discussion of the implications of these assumptions in cosmography see Chap. 8 of \cite{Ellis2}). However, although there are efforts to derive $\Lambda$ from non-cosmological physics, it remains a fitting parameter of the model. The appropriate cosmographical framework was put forward in the '80s by G.F.R. Ellis, R. Maartens, W. Stoeger, and A. Whitman \cite{EllisPhRep} (see also \cite{Ellis2}) by characterizing the  set of cosmological observables on the past lightcone which, together with the Einstein field equations, allows to reconstruct the spacetime geometry in a way adapted to the process of observation 
\cite{EllisPhRep}, \cite{Choquet0}, \cite{Choquet1}.     \\  
\\
In this paper we address an important step in this cosmographical framework. In particular we discuss a rigorous procedure for quantifying the difference between our past lightcone and the reference past lightcone that, for consistency, we associate with the fiducial large-scale FLRW  spacetime.  This result is made possible by exploiting the scale-dependent (harmonic map type) distance functional between past lightcones recently introduced by us in \cite{CarFam}, and which extended the light-cone theorem \cite{ChoquetLightCone}. We  express this functional  in terms of the physical quantities that characterize  measurements along our past lightcone, namely the area distance and the lensing distortion,  also briefly addressing the very delicate problem of  the presence of  lightcone caustics.  This analysis works beautifully and seems to remove several of the difficulties encountered in comparing the actual geometry of our past lightcone with the geometry of a fiducial FLRW lightcone of choice.  We also discuss how, from the point of view of the FLRW geometry,  this distance functional may be interpreted as a  scale-dependent effective field that may be of relevance in selecting the FLRW model that best fits the observative data. In this connection and in line with the introductory remarks above its worthwhile to stress that our choice of a reference FLRW spacetime is strictly related to the prevalence of this family of metrics in discussing the $\Lambda\mathrm{CDM}$ model. The results presented here can be easily extended to more general reference metrics. It is also important to make clear that in this paper we are not addressing the extremely delicate averaging problem on the past lightcone, a problem to which
Maurizio Gasperini has significantly contributed with the seminal paper \cite{Gasperini}, and that has seen importat recent progress in \cite{BuchertLC}... \emph{but the past lighcone routes are still tough and the landscape rugged ...} .
\section{The past light cone and the celestial sphere}
\label{sect2}
Throughout this paper  $(M, g)$ denotes a  cosmological spacetime  where $g$ is a Lorentzian metric, and where $M$ is  a smooth $4$-dimensional manifold which for our purposes we can assume diffeomorphic to $\mathbb{R}^4$ (or to $V^3\times\mathbb{R}$, for some smooth compact or complete $3$--manifold $V^3$).  In local coordinates $\{x^i\}_{i=1}^4$, we write $g=g_{ik}dx^i\otimes dx^k$, where the metric components $g_{ik}\,:=\,g(\partial_i, \partial_k)$
 in the coordinate basis $\{\partial_{i}:=\partial /\partial  x^i\}_{i=1}^4$ have the Lorentzian signature $(+,+,+,-)$, and the Einstein summation convention is in effect\footnote{If not otherwise stated we adopt geometrical units, $c\,=\,1\,=\,G$.}. We assume that $(M, g)$  is associated with the evolution of a universe which is (statistically) isotropic and homogeneous on sufficiently large scales
 $L\,>\,L_0$ where, according to the introductory remarks, we indicatively assume $L_0\,\cong\, 100h^{-1}\; \mathrm{Mpc}$, and let local inhomogeneities dominate for $L\,<\,L_0$.   The matter content in $(M, g)$ is phenomenologically described by a (multi-component)  energy-momentum tensor $T\,=\,T_{ik}\,dx^i\otimes dx^k$, (typically  in the form of a perfect fluid, dust, and radiation). If not otherwise stated, the explicit expression of $T$ is not needed for our analysis. We assume that in $(M, g)$ the  motion of the matter components  characterize a \emph{phenomenological Hubble flow} that generates a family of preferred world-lines parametrized by proper time $\tau$
\begin{eqnarray}
\label{observers}
\gamma_s\,:\,\mathbb{R}_{>0}\,&\longrightarrow & \,(M, g)\\
\tau\,&\longmapsto&\,\gamma_s(\tau)\;,\nonumber
\end{eqnarray}
 and labeled by suitable comoving (Lagrangian) coordinates  $s$ adapted to the  flow.
 We denote by $\dot{\gamma}_s\,:=\,\frac{d\gamma_s(\tau)}{d\tau}$,\,\,$g(\dot\gamma_s,  \dot\gamma_s)\,=\,-1$, the corresponding  $4$-velocity field.  For simplicity, we assume that at the present era these worldlines  are geodesics, \textit{i.e.} $\nabla_{\dot\gamma_s}\,\dot\gamma_s\,=\,0$. This phenomenological  Hubble flow is strongly affected by the peculiar motion of the astrophysical sources and by the complex spacetime geometry that dominates on the pre-homogeneity scales. In particular, it exhibits a complex pattern of fluctuations with respect to the linear \emph{FLRW Hubble flow} that sets in, relatively to the standard of rest provided by the cosmic microwave background (CMB), when we probe the homogeneity scales, $L\,\gtrsim\, 100h^{-1}\; \mathrm{Mpc}$. Again, we stress that the transitional region between the phenomenological Hubble flow and the statistical onset of the large-scale FLRW linear Hubble flow is quite uncertain and still actively debated \cite{Wiltshire}.  If we adopt the weak form of the cosmological principle described in the introduction, $(M, g, \gamma_s)$ can be identified with the phenomenological background spacetime or \textit{Phenomenological Background Solution (PBS)}\cite{KolbMarraMatarrese} associated with the actual cosmological data gathered from our past lightcone observations. In the same vein, we define \textit{Phenomenological Observers}  the collection of observers $\{\gamma_s\}$ comoving with the phenomenological Hubble flow (\ref{observers}). Since in our  analysis we fix our attention on  a given observer, we drop the subscript $s$ in (\ref{observers}), and  describe a finite portion of  the observer's world-line with the timelike geodesic segment  $\tau\,\longmapsto\,\gamma(\tau)$,\,$-\delta<\tau<\delta$, \, for some $\delta>0$, \,\,where  $p\,:=\,\gamma(\tau=0)$ is  the selected event corresponding to which the cosmological data are gathered. To organize and describe these data in the local rest frame of the observer $p\,:=\,\gamma(\tau=0)$, let $\left(T_p M,\,g_p,\,\{E_{(i)}\}\right)$ be the tangent space to $M$ at $p$ endowed with a $g$-orthonormal frame $\{E_{(i)}\}_{i=1,\ldots,4}$,\,\,$g_p\left(E_{(i)}, E_{(k)}\right)=\eta_{ik}$, where $\eta_{ik}$ is the Minkowski metric, and where we identify $E_{(4)}$ with the observer $4$-velocity $\dot{\gamma}(\tau)|_{\tau=0}$,\;\,\emph{i.e.}\,\,  $E_{(4)}\,:=\,\dot{\gamma}(\tau)|_{\tau=0}$. Thus, if we denote by $\{\breve{E}^{\,(i)}\}_{i=1,\ldots,4}$, the $1$-forms basis dual to $\{E_{(i)}\}_{i=1,\ldots,4}$, we write
\begin{eqnarray}
g_p\,=\,\eta_{ik}\,\breve{E}^{\,(i)}\,\otimes\,\breve{E}^{\,(k)}\,.
\end{eqnarray}
Since we have the distinguished choice $E_{(4)}\,:=\,\dot{\gamma}(\tau)|_{\tau=0}$ for the timelike  basis vector $E_{(4)}$, we can also introduce in $\left(T_pM,\,\{E_{(i)}\}\right)$ a reference positive definite metric $g^{(\delta)}_p$ associated with the frame $\{E_{(i)}\}_{i=1,\ldots,4}$  by setting
\begin{eqnarray}
\label{Riemp}
g^{(\delta)}_p\,:=\,\delta_{ik}\,\breve{E}^{\,(i)}\,\otimes\,\breve{E}^{\,(k)}\,,
\end{eqnarray}
where $\delta_{ik}$ denote the components of the standard Euclidean metric. As discussed in detail by Chen and LeFloch \cite{LeFloch},\,  this reference metric comes in handy in the characterization of the functional Lipschitz and Banach space norms of tensor fields defined on the past lightcone\footnote{The indefinite character of a Lorentzian metric makes it unsuitable for defining integral norms of tensor fields, and for such a purpose one is forced to introduce a reference positive definite metric. In particular, by exploiting the Nash embedding theorem, one typically uses the Euclidean metric and the associated definitions of the functional space of choice, say a Sobolev space of tensor fields. Different choices of reference metrics, as long as they are of controlled geometry, induce equivalent Banach space norms. In our case, we can exploit the natural choice provided by (\ref{Riemp}) by using normal coordinates and identifying $(T_pM,\,\{E_{(i)}\}, \,g_p^{\delta})$ with the Euclidean space $(\mathbb{R}^4,\,g_p^{\delta})$.}.

\subsection{The celestial sphere}
\label{celestialS}     
Let
\begin{equation}
\label{MinkLightCone}
C^-\left(T_pM,\,\{E_{(i)}\} \right)\,:=\,\left\{X\,=\,\mathbb{X}^iE_{(i)}\,\not=\,0\,\in\,T_pM\,\,|\,g_p(X, X)\,=\,0,\,\,\mathbb{X}^4+r=0 \right\}\;,
\end{equation}
\begin{equation}
\label{pastMinkLightCone}
\overline{C^-\left(T_pM,\,\{E_{(i)}\} \right)}\,:=\,\left\{X\,=\,\mathbb{X}^iE_{(i)}\,\not=\,0\,\in\,T_pM\,\,|\,
g_p(X, X)\,\leq\,0,\,\,\mathbb{X}^4+r\,\leq\,0 \right\}\;,
\end{equation}
respectively denote the set of past-directed null vectors and the set of  past-directed causal vectors in $(T_pM,\,\{E_{(i)}\})$, where 
\begin{equation}
\label{radial}
r:=(\sum_{a=1}^3(\mathbb{X}^a)^2)^{1/2}\,,
\end{equation}
is the radial coordinate in  the hyperplane 
$\mathbb{X}^4\,=\,0\,\subset\,T_pM$ parametrizing the one-parameter family of $2$-spheres 
\begin{equation}
\label{celestialR}
 \mathbb{S}^2_r(T_pM)\,:=\,\{X\in C^-\left(T_pM, \{E_{(i)}\} \right)\,|\,\, \mathbb{X}^4\,=\,-\,r,\,\,\,\sum_{a=1}^3(\mathbb{X}^a)^2=r^2,\,\,r\in\,\mathbb{R}_{> 0} \}\,,
 \end{equation}
that foliates $C^-\left(T_pM, \{E_{(i)}\}\right)/\{p\}$. The sphere $\mathbb{S}^2_r(T_pM)$ can be thought of as providing a representation of the sky directions, at a given value of $r$, in the rest space $\left(T_pM, \{E_{(i)}\} \right)$ of the (instantaneous) observer $(p,\dot\gamma(0))$. In particular, the 2-sphere   $\left.\mathbb{S}^2_r(T_pM)\right|_{r=1}$ or, equivalently, its projection on the hyperplane 
$\mathbb{X}^4\,=\,0$ in $T_pM$, 
\begin{equation}
\label{celestialS}
\mathbb{S}^2\left(T_pM\right)\,:=\,\left\{X\,=\,\mathbb{X}^iE_{(i)}\,\not=\,0\,\in\,T_pM\,\,|\,\,\mathbb{X}^4=0,\,\,\sum_{a=1}^3(\mathbb{X}^a)^2=1 \right\}\;,
\end{equation}
can be used to parametrize the (spatial) past directions of sight constituting the field of vision of the observer $(p,\, \dot\gamma(0))$. In the sense described by R. Penrose  \cite{RindPen},  this is a representation of 
the abstract sphere $\mathcal{S}^-(p)$ of past null directions parameterizing the past-directed null geodesics through $p$. Explicitly, let
\begin{eqnarray}
{n}(\theta, \phi)\,&:=&\,\sum_{a=1}^3\,{n}^a(\theta, \phi)\,{E}_{(a)}\\
&=&\,\cos\phi\sin\theta\, {E}_{(1)}\,+\,\sin\phi\sin\theta\, {E}_{(2)}\,+\,cos\,\theta\, {E}_{(3)}\,,\,\,\,\,
0\leq\theta\leq\pi,\,\,0\leq\phi<2\pi\,,\nonumber
\end{eqnarray}
denote the spatial direction in $T_pM$ associated with the point $(\theta, \phi)\,\in\,\mathbb{S}^2\left(T_pM\right)$, (by abusing notation, we often write ${n}(\theta, \phi)\,\in\,\mathbb{S}^2\left(T_pM\right)$). Any such spatial direction characterizes a corresponding past-directed null vector $\ell(\theta, \phi)\,\in\,\left(T_pM, \{E_{(i)}\} \right)$,  
\begin{equation}
\label{elle}
\ell(\theta, \phi)\,=\,\left(n(\theta,\,\phi),\,-\,\dot{\gamma}(\tau)|_{\tau=0}   \right)\,=\,\sum_{a=1}^3\, n^a(\theta, \phi) E_{(a)}\,-\,E_{(4)}\,,
\end{equation}
 normalized according to
\begin{equation}
g_p\left(\ell(\theta, \phi),\dot{\gamma}(\tau)|_{\tau=0}\right)\,=\,g_p\left(\ell(\theta, \phi),E_{(4)}\right)\,=\,1\,.
\end{equation}
The corresponding past-directed null rays 
\begin{equation}
\mathbb{R}_{\geq 0}\,\ni\,r\,\longmapsto\,r\,\ell({n}(\theta, \phi))\,,\,\,\,\,\,\,\,\,(\theta, \phi)\,\in\,\mathbb{S}^2\left(T_pM\right)\,,
\end{equation}
generate $C^-\left(T_pM, \{E_{(i)}\} \right)$.
Note that in such a kinematical setup for the instantaneous rest space $\left(T_pM,\,\{E_{(a)}\}\right)$  of the observer $(p,\, \dot\gamma(0))$,  a photon reaching $p$ from the past-directed null direction $\ell(\theta, \phi)$, is characterized  by  the (future-pointing) wave vector 
\begin{equation}
k(\theta, \phi)\,:=\,-\,{\nu}\,\ell(\theta, \phi)\,\in\, T_pM\,,
\end{equation}
where $\nu\,=\,-\,g_p\left(k,\, E_{(4)}\right)$ is the photon frequency as measured by the instantaneous observer ${\gamma}(\tau)|_{\tau=0}$. The spherical surface $\mathbb{S}^2\left(T_pM\right)$ endowed with the standard round metric 
\begin{equation}
\label{roundmet}
\widetilde{h}(\mathbb{S}^2)\,=\,d\theta^2\,+\,\sin^2\theta\,d\phi^2\,,
\end{equation}
and the associated area form
$d\mu_{\mathbb{S}^2}\,=\,\sqrt{\det(\widetilde{h}(\mathbb{S}^2))}\,d\theta d\phi\,=\,\sin\theta\,d\theta d\phi$,
defines \cite{RindPen} the \emph{celestial sphere} 
\begin{equation}
\label{celstSphere}
\mathbb{C\,S}(p)\,:=\,\left(\mathbb{S}^2\left(T_pM\right),\,\widetilde{h}(\mathbb{S}^2)     \right)\,
\end{equation}
providing, in the instantaneous rest space $\left(T_pM, \{E_{(i)}\} \right)$, the geometrical representation of the set of all directions towards which the observer can look at astrophysical sources from her instantaneous location in $(M, g)$. In this connection, 
$d\mu_{\mathbb{S}^2}$ can be interpreted as the element of solid angle subtended on the celestial sphere $\mathbb{C\,S}(p)$ by the observed astrophysical sources. It is also useful to keep track of the radial coordinate\footnote{To avoid any misunderstanding we stress that $r$ is not a distance parameter on the past light cone with vertex in $p\in (M, g)$.} $r$ as a possible parametrization of the past-directed null geodesics, and introduce a celestial sphere that provides also this information according to
\begin{equation}
\label{celstSphereR}
\mathbb{C\,S}_r(p)\,:=\,\left(\mathbb{S}^2_r\left(T_pM\right),\,r^2\widetilde{h}(\mathbb{S}^2) \right)\,.
\end{equation}
Lacking a better name, we shall refer to $\mathbb{C\,S}_r(p)$ as the \emph{celestial sphere at radius $r$} in $\left(T_pM, \{E_{(i)}\} \right)$.  
The celestial sphere $\mathbb{C\,S}(p)$ plays a basic role in what follows since it provides the logbook where astrophysical data are recorded.\\
 Let $m_{(\alpha)}(\theta, \phi)\in\,T_pM$, with $\alpha\,=\,2,3$, denote  two  spatial $g_p$-orthonormal vectors  spanning the tangent space  $T_{(\theta, \phi)}\mathbb{S}^2\left(T_pM\right)$ to $\mathbb{S}^2\left(T_pM\right)$ at the point $(\theta, \phi)$, \emph{i.e.},  
\begin{equation}
g_p\left(m_{(\alpha)}, n \right)\,=\,0\,=\,g_p\left(m_{(\alpha)}, E_{(4)}\right),\,\,g_p\left(m_{(\alpha)}, m_{(\beta)}\right)\,=\,\delta_{\alpha\beta}\,.
\end{equation}
The tetrad  
\begin{equation}
\label{SachsBasis}
\left(n, m_{(2)}, m_{(3)}, \ell(n) \right)
\end{equation}
provides a basis for $T_pM$ (the Sachs basis), and the pair $\left(T_{(\theta, \phi)}\mathbb{S}^2\left(T_pM\right),\;m_{(\alpha)}(\theta, \phi)\right)$ defines 
the \emph{screen plane} $T_{n}\mathbb{C\,S}(p)$ associated with the direction of sight ${n}(\theta, \phi)\in\,\mathbb{C\,S}(p)$ in the celestial sphere $\mathbb{C\,S}(p)$, \emph{i.e.}
\begin{equation}
\label{screenM}
T_{n}\mathbb{C\,S}(p)\,:=\,\left(T_{(\theta, \phi)}\mathbb{S}^2\left(T_pM\right),\;m_{(\alpha)}(\theta, \phi)\right)\,.
\end{equation}
In the instantaneous rest space  of the observer, the screen $T_{(\theta,\phi)}\mathbb{C\,S}(p)$ is the (spatial) $2$-plane on which the apparent image of the astrophysical source, pointed by the direction $n\in \mathbb{C\,S}(p)$, is by convention displayed.

\subsection{Sky sections and observational coordinates on the past light cone}
\label{subsect2}
We transfer the above kinematical setup from $T_pM$ to $(M, g)$  by using the exponential map based at $p$,
\begin{eqnarray}
\label{expmapdef}
\exp_p\,:\,W_p\,\subseteq \,T_pM\,&\longrightarrow&\;\;\;\;M\\
X\;\;\;\;\;\;&\longmapsto&\;\;\;\;exp_p\,(X)\,:=\,\lambda_X(1)\,,\nonumber
\end{eqnarray}
where $\lambda_X\,:\,I_W\,\longrightarrow\,(M, g)$, for some maximal interval $I_W\subseteq\mathbb{R}_{\geq0}$,  is the past-directed causal geodesic emanating from the point $p$ with initial tangent vector $\dot{\lambda}_X(0)\,=\, X\in W_p$, and where 
$W_p\,\subseteq \,T_pM$ is the maximal domain of $\exp_p$. Thus, the past lightcone $\mathscr{C}^-(p,g)\,\in\,(M, g)$ with the vertex  at $p$,\,\textit{i.e.} the set of all events $q\in (M, g)$ that can be reached from $p$ along the past-pointing null geodesics $r\,\longmapsto\,\exp_p(r \ell(n(\theta, \phi)))$,\,$r\in\,I_W$, $(\theta, \phi)\in\mathbb{C\,S}(p)$, can be represented as
\begin{equation}
\label{pastcone2}
\mathscr{C}^-(p,g)\,:=\,\exp_p\left[W_p\cap C^-\left(T_pM, g_p \right)\right]\,,
\end{equation}
and  the portion of $\mathscr{C}^-(p,g)$ accessible to observations for a given value $r_0\,\in\,I_W$ of the affine parameter $r$ is given by
\begin{equation}
\label{pastcone3}
\mathscr{C}^-(p,g;\,r_0)\,:=\,\left\{\left. q\,\in\,M\,\right|\,\,q\,=\,\exp_p(r \ell(n(\theta, \phi))),\, \, 0\leq r < r_0,\,\, (\theta, \phi)\in\mathbb{C\,S}(p)  \right\}\nonumber\,.
\end{equation}
The exponential map representation, on the celestial spheres $\mathbb{C\,S}(p)$ and $\mathbb{C\,S}_r(p)$, provides a natural setup for  a description of observational data gathered from $\mathscr{C}^-(p,g)$. It emphasizes the basic role of past-directed null geodesics and provides the framework for interpreting the physical data in the local rest frame of the observer at $p$. In particular, it allows us to represent on  
$\mathbb{C\,S}(p)$ and $\mathbb{C\,S}_r(p)$ the actual geometry of the observed sky at a given length scale. This role is quite effective in a neighborhood of $p$, where we can introduce normal coordinates associated with $\exp_p$, but it is delicate to handle in regions where $\exp_p$ is not a diffeomorphism of $W_p\cap C^-\left(T_pM, g_p \right)$  onto its image. 
To set notation, our strategy is to start  with the standard description \cite{EllisPhRep}, \cite{Ellis2} of observational coordinates on $\mathscr{C}^-(p,g)$ associated with the usual assumption that the exponential map is a diffeomorphism\footnote{From an observational point of view, this is the  geometrical set-up proper of the \emph{weak lensing regime} describing the alteration, due to the  effect of gravity, of  the apparent shape and brightness of astrophysical sources.} 
in a sufficiently small neighborhood of $p$, and then we move to the more general, low regularity, Lipschitz case. In this connection, it is worthwhile to stress that the standard normal coordinates description is strictly associated with the assumption that the metric of $(M, g)$ is sufficiently regular, with components $g_{ij}(x^\ell)$ which are at least twice continuously differentiable, \emph{i.e.}\, $g_{ij}(x^\ell)\,\in\,C^k(\mathbb{R}^4, \mathbb{R})$, \, for $k\geq 2$. Under this hypothesis, there is  a star-shaped neighborhood $N_0(g)$ of  $0$ in $W_p\subseteq T_pM$ and a corresponding geodesically convex neighborhood of $p$, $U_p 	\subseteq\,(M, g)$, restricted to which $\exp_p\,:\,N_0\,\subseteq \,T_pM\,\longrightarrow\;U_p\,\subseteq\,M$ is a diffeomorphism. In such $U_p$ we can  introduce geodesic normal coordinates $(x^i)$ according to
\begin{eqnarray}
\label{geodnormal}
x^i\,:=\,\mathbb{X}^i\,\circ \,\exp_p^{-1}\,:\,M\cap\,U_p\,&\longrightarrow&\,\mathbb{R}^4\\
q\,\,\,\,\,\,\,&\longmapsto&\,x^i(q)\,:=\, \mathbb{X}^i\left(\exp_p^{-1}(q)\right)\nonumber
\end{eqnarray}
where $\mathbb{X}^i\left(\exp_p^{-1}(q)\right)$ are the components, in the $g$-orthonormal frame $\{E_{(i)}\}$, (or with respect to the corresponding  basis (\ref{SachsBasis})), of the vector $\exp_p^{-1}(q)\,\in\,W_p\subseteq T_pM$. Thus, in $\mathscr{C}^-(p,g)\cap\,U_p$  we can write,  
\begin{eqnarray}
\label{exp1}
\exp_p\
\,:\,C^-\left(T_pM,\,\{E_{(i)} \}\right)\cap\,N_0(g)\,&\longrightarrow&\,\mathscr{C}^-(p,g)\,\cap\,U_p\\
r \ell(n(\theta, \phi))\,=\,r \left(n^a(\theta, \phi) E_{(a)}\,-\,E_{(4)}\right)\,\,\,\,\,\,\,\,\,\,&\longmapsto&\,\exp_p(r \ell(n))\,=\,q\nonumber\\
\Longrightarrow\,q&\longmapsto&\,\{x^i(q)\,:=\,\exp_p^{-1}(q)\,=\,\left(r\,n^a(\theta,\phi),\,-\,r\right)\}\,\nonumber\;.
\end{eqnarray}
According to (\ref{pastcone2}) and to the Gauss lemma applied to $\exp_p\,:\,C^-\left(T_pM,\, \{E_{(i)} \}\right)\cap\,N_0(g)\,\longrightarrow\,\mathscr{C}^-(p,g)\,\cap\,U_p$, the past ligh cone region
$\mathscr{C}^-(p, g)\,\cap\,U_p\setminus \{p\}$ is foliated by the $r$-dependent family of   2--dimensional surfaces $\Sigma(p, r)$, the \emph{cosmological sky sections}, defined by
\begin{equation}
\label{sigmapr}
\Sigma(p, r)\,:=\,\exp_p\left[\mathbb{C\,S}_r(p) \right]\,=\, \left\{\left.\exp_p\left(r\,\ell({n}(\theta, \phi))\right)\,\right|\,\, (\theta, \phi) \,\in\,\mathbb{C\,S}(p)\right\}\,,
\end{equation}
and $g$-orthogonal to all null geodesics originating at $p$, \emph{i.e.} 
\begin{equation}
\label{GaussLemma}
\left.g\left(d_{(r,\theta,\phi)}\exp_p(\ell(r, \underline{n})),\,d_{(r,\theta,\phi)}\exp_p(\underline{v})\right)\right|_{\exp_p(\ell(r, \underline{n}))}\,=\,0\,.
\end{equation}
Here $d_{(r,\theta,\phi)}\exp_p(...)$ denotes the tangent mapping associated to $\exp_p$ evaluated at the point $(\theta,\phi)\,\in\,\mathbb{S}^2_r(p)$, and  $\underline{v}\,\in\,T_{\theta, \phi}\,\mathbb{S}^2_r(p)$ is the generic vector tangent to $\mathbb{S}^2_r(p)$. In $\mathscr{C}^-(p, g)\,\cap\,U_p\setminus \{p\}$,\,
each surface $\Sigma(p, r)$ is topologically a 2-sphere endowed with the $r$-dependent two-dimensional Riemannian metric
\begin{equation}
\label{Sigmametric}
g|_{\Sigma(p, r)}\,:=\,\iota_r^*\,\left.g\right|_{\mathscr{C}^-(p, g)}
\end{equation}
induced by the inclusion $\iota_r:\Sigma(p, r)\,\hookrightarrow\,\mathscr{C}^-(p, g)$ of $\Sigma(p, r)$  into $\mathscr{C}^-(p, g)\,\cap\,U_p\setminus \{p\}$. We can pull back this metric to the celestial sphere $\mathbb{C\,S}_r(p):=\,\left(\mathbb{S}^2_r\left(T_pM\right),\,r^2\widetilde{h}(\mathbb{S}^2) \right)$ by using the exponential map according to
\begin{equation}
\label{metrich}
h(r,\theta, \phi)\,:=\,\left.\left(\exp_p^*\,g|_{\Sigma(p, r)}\right)_{\alpha\beta}\,dx^\alpha dx^\beta\right|_{r},\,\,\,\,\alpha,\,\beta\,=\,2, 3,\,\,\,\,\,\,x^2:=\theta,\,x^3:=\phi\,.
\end{equation}
This metric can be profitably compared with the pre-existing round metric $r^2\widetilde{h}(\mathbb{S}^2)$ on  $\mathbb{C\,S}_r(p)$ (see (\ref{roundmet}) and (\ref{celstSphereR})). To this end, let $r\,n(\theta, \phi)\,\in\,\mathbb{C\,S}_r(p)$ be the direction of sight pointing, in the celestial sphere $\mathbb{C\,S}_r(p)$, to the (extended) astrophysical source located around the point $q\,\in\,\Sigma(p, r)$. If 
$r \ell(n(\theta, \phi))\,=\,r \left(n^a(\theta, \phi) E_{(a)}\,-\,E_{(4)}\right)$ is the corresponding null direction in $C^-\left(T_pM,\,\{E_{(i)} \}\right)$, then according to
(\ref{exp1}) we have $\exp_p(r \ell(n))\,=\,q$ and, via the exponential map along the 
past-directed null geodesic reaching the observer located at $p$ from the astrophysical source located at $q$, we can pull-back the area element of $\left(\Sigma(p,r),\,g|_{\Sigma(p, r)}\right)$ on the celestial sphere $\mathbb{C\,S}_r(p)$ of the observer at $p$. We have
\begin{equation}
\label{hvolume}
d\mu_{h(r)}(p,n(\theta, \phi), r)\,:=\,\exp_p^*d\mu_{g|_{\Sigma(p, r)}}\,\circ\,\exp_p(r \ell(n))\,=\,\sqrt{\det(h(r,\theta,\phi))}\,d\theta d\phi\,.
\end{equation}     
This defines the area element associated with the metric (\ref{metrich}), and  can be interpreted \cite{Ellis2} as the cross-sectional area element at the source location as seen by the observer at $p$.
Since the round measure $d\mu_{\mathbb{S}^2_r}=r^2\,d\mu_{\mathbb{S}^2}=r^2\,\sin\theta\,d\theta\,d\varphi$ and the actual physical measure $d\mu_{h(r)}$ are both defined over the celestial sphere $\mathbb{C\,S}_r(p)\in\,T_pM$, we  can introduce the relative density of $d\mu_{h(r)}$ with respect to the Euclidean solid angle measure $d\mu_{\mathbb{S}^2}$, \emph{viz.} the function $D(r, \theta, \phi)$  defined by the relation 
\begin{equation}
\label{Di}
d\mu_{h(r)}\,  =\,  D^2(r, \theta, \phi)\,d\mu_{\mathbb{S}^2}\,,
\end{equation}
or equivalently, $\sqrt{\det(h(r,\theta,\phi))}\,=\,D^2(r, \theta, \phi)\,\sqrt{\det(\widetilde{h}(\mathbb{S}^2))}$. The function $D(r, \theta, \phi)$  is \emph{the  observer area distance} \cite{EllisPhRep}, \cite{Ellis2}, \cite{Hogg1}. By definition, it provides the ratio of an object's  cross sectional area to its (apparent) angular size as seen on the celestial sphere $\mathbb{S}^2(p)\,\subset\,T_pM$. Roughly speaking, it converts the angular separations as seen in the images of an astrophysical source, gathered by the observer at $p$, into proper separations at the source.  In general, 
$D(r)\,:=\,\left.D(r, \theta, \phi)\right|_{\theta, \phi= const.}$ cannot be used as an affine parameter along the past-directed null geodesic $r\,\mapsto\,\exp_p(k(r, \underline{n}))$ since  it is not a monotonic function of $r$, (for instance in FLRW models, monotonicity fails around $z\,\sim\,1$).  However, if  we have an accurate knowledge of the brightness and of the spectrum of the astrophysical source seen at the past light cone location $q\,:=\,\exp_p(\ell(r, \underline{n}))\,\in\,\mathscr{C}^-(p, g)$, then  $D(r, \theta, \phi)$ is, at least in principle, a measurable quantity  (see paragraph 4.3 of \cite{EllisPhRep} and 7.4.3 of \cite{Ellis2} for a discussion of this point\footnote{Beware that in \cite{EllisPhRep}, the observer area distance $D^2(r, \theta, \phi)$ is denoted by $r$, whereas our $r$ corresponds to their $y$.}). As stressed above, we can also compare  the physical metric (\ref{metrich}), $h(r,\theta, \phi)\,:=\,\left.\left(\exp_p^*\,g|_{\Sigma(p, r)}\right)_{\alpha\beta}\,dx^\alpha dx^\beta\right|_{r}$, \, with the round metric $r^2\widetilde{h}(\mathbb{S}^2)$ pre-existing on the celestial sphere $\mathbb{C\,S}_r(p)$, and   introduce \cite{EllisPhRep}, \cite{Ellis2} the set of  functions $\mathcal{L}_{\alpha\beta}(r, \theta, \phi)$, $\alpha,\,\beta\,=\,2,3$, implicitly defined by representing (\ref{metrich}) in the distorted polar form 
\begin{equation}
\label{etametric0}
\left.h_{\alpha\beta}\right|_{\mathbb{S}^2_r}\,=\, D^2(r, \theta, \phi)\left(\widetilde{h}_{\alpha\beta}(\mathbb{S}^2)\,+\,\mathcal{L}_{\alpha\beta} \right)\;. 
\end{equation}
We normalize this representation  by  imposing \cite{EllisPhRep}  that, in the limit $r\,\searrow\, 0$, the distortion,\, $\mathcal{L}_{\alpha\beta}(r,\theta, \phi)=\frac{h_{\alpha\beta}(r,\theta, \phi)}{D^2(r, \theta\,\phi}-\widetilde{h}_{\alpha\beta}(\mathbb{S}^2)$,\, of the normalized metric $h(r)/D^2(r)$ with respect to the round metric $\widetilde{h}(\mathbb{S}^2)$ goes to zero uniformly,\, \textit{i.e.},
\begin{equation}
\label{limitr1}
\left.\lim_{r\searrow 0}\right|_{x^4=0}\,\frac{h_{\alpha\beta}(r, \theta,\phi)\,dx^\alpha dx^\beta}{D^2(r, \theta,\phi)}\,
=\,d\theta^2\,+\,\sin^2\theta\,d\phi^2\;.
\end{equation}
From the relation $D^{-\,2}\,h_{\alpha\beta}\,=\,\widetilde{h}_{\alpha\beta}(\mathbb{S}^2)\,+\,\mathcal{L}_{\alpha\beta}$ we also compute
\begin{equation}
\label{deterL}
D^{-\,2}\,\widetilde{h}^{\mu\beta}\,h_{\alpha\beta}=\,
\delta_{\alpha}^\mu\,+\,\mathcal{L}_{\alpha}^\mu\,\Longrightarrow\,\det
\left(\delta_{\alpha}^\mu\,+\,\mathcal{L}_{\alpha}^\mu \right)\,=\,1\,,
\end{equation} 
where, for rising indexes, we used the inverse round metric $\widetilde{h}^{\mu\beta}(\mathbb{S}^2)$ to write  $\mathcal{L}_{\alpha}^\mu :=\,\widetilde{h}^{\mu\beta}(\mathbb{S}^2)\,\mathcal{L}_{\alpha\beta}$, and where we have exploited the relation $\det\left({\widetilde{h}^{\mu\beta}\,h_{\alpha\beta}}\right)\,=\,D^4$, direct consequence of ${\det(h)}\,=\,D^4\,\det(\widetilde{h}(\mathbb{S}^2))$ (see (\ref{Di})). Since
\begin{equation}
\,\det
\left(\delta_{\alpha}^\mu\,+\,\mathcal{L}_{\alpha}^\mu \right)\,=\,1\,+\,\mathrm{tr}_{\mathbb{S}^2}\left(\mathcal{L}_{\alpha}^\mu\right)\,+\,\det\left(\mathcal{L}_{\alpha}^\mu \right)\,,
\end{equation} 
from relation (\ref{deterL}) it follows that
\begin{equation}
\label{notracefree}
\mathrm{tr}_{\mathbb{S}^2}\left(\mathcal{L}_{\alpha}^\mu\right)\,+\,\det\left(\mathcal{L}_{\alpha}^\mu \right)\,=\,0\,,
\end{equation}
which implies that $\mathcal{L}_{\alpha}^\mu$ cannot be trace-free. Roughly speaking, $\mathcal{L}_{\alpha\beta}(r)$ can be interpreted as the image distortion of  the sources on $\left(\Sigma(p, r), h(r)\right)$ as seen by the observer at $p$ on her celestial sphere. It can in principle be directly observed and it can be related to the gravitational lensing shear \cite{EllisPhRep}, (see also chap. 8 of \cite{Ellis2}). Explicitly, let us compute the deformation tensor $\Theta_{\alpha\beta}$ defined by the rate of variation of the metric tensor $h(r)$ as $r$ varies. Dropping the angular dependence for notational ease, we get
\begin{eqnarray}
\label{Distortion}
\Theta_{\alpha\beta}\,:=\,
\frac{d}{d r}\,h_{\alpha\beta}(r)\,&=&\,\frac{d}{d r}\,\left[D^2(r)\left(\widetilde{h}_{\alpha\beta}(\mathbb{S}^2)\,+\,\mathcal{L}_{\alpha\beta}(r) \right)  \right]\\
&=&\,2h_{\alpha\beta}(r)\,\frac{d}{d r}\,\ln D(r)\,+\,D^2(r)\,\frac{d}{d r}\,\mathcal{L}_{\alpha\beta}(r)\,,  \nonumber
\end{eqnarray}
where we exploited $d \widetilde{h}_{\alpha\beta}(\mathbb{S}^2)/dr\,=\,0$ and rewrote $D(r)dD(r)/dr$ as $D^2(r)d\ln\,D(r)/dr$. Similarly, from the defining relation $\sqrt{\det(h(r,\theta,\phi))}\,=\,D^2(r, \theta, \phi)\,\sqrt{\det(\widetilde{h}(\mathbb{S}^2))}$, (see (\ref{Di})), we compute
\begin{eqnarray}
\frac{d}{d r}\,\sqrt{\det(h(r))}\,&=&\,\frac{d}{d r}\,\left(D^2(r)\,\sqrt{\det(\widetilde{h}(\mathbb{S}^2))}  \right)\,=\,2\,\sqrt{\det(h(r))}\,\frac{d}{d r}\,\ln D(r)\\
\Rightarrow\,\,\frac{d}{d r}\,\ln\sqrt{\det(h(r))}\,&=&\,2\,\frac{d}{d r}\,\ln D(r)\nonumber
\,.
\end{eqnarray}
Inserting this relation in (\ref{Distortion}) we obtain
\begin{equation}
\Theta_{\alpha\beta}\,:=\,h_{\alpha\beta}(r)\,\frac{d}{d r}\,\ln\sqrt{\det(h(r))}\,+\,D^2(r)\,\frac{d}{d r}\,\mathcal{L}_{\alpha\beta}(r)\,.
\end{equation}
The shear $\widetilde{\sigma}_{\alpha\beta}$ is the trace-free part of this expression, $\widetilde{\sigma}_{\alpha\beta}:=\Theta_{\alpha\beta}-\frac{1}{2}h_{\alpha\beta}\,h^{\mu\nu}\Theta_{\mu\nu}$. Since 
\begin{equation}
\frac{1}{2}h_{\alpha\beta}\,h^{\mu\nu}\Theta_{\mu\nu}\,=\,\frac{1}{2}h_{\alpha\beta}\,h^{\mu\nu}\frac{d}{d r}h_{\mu\nu}\,=\,h_{\alpha\beta}\,\frac{d}{d r}\,\ln\sqrt{\det(h(r))}\,,
\end{equation}
we eventually get
\begin{equation}
\widetilde{\sigma}_{\alpha\beta}\,=\,D^2(r)\,\frac{d \mathcal{L}_{\alpha\beta}(r)}{d r}\;,
\end{equation}
as might have been expected. Note that, in contrast to $\mathcal{L}_{\alpha\beta}$,  $\widetilde{\sigma}_{\alpha\beta}$ is trace-free (but with respect to the physical metric $h_{\alpha\beta}$). Now, let us introduce the other basic player of our narrative.

\section{The background FLRW past light cone.}   

As already pointed out, the standard $\Lambda$CDM model is built on the assumption that over scales $L\,>\,100\,h^{-1}\,\mathrm{Mpc}$, the phenomenological background spacetime $(M, g, \gamma_s)$ follows on average the dynamics of a FLRW model with a (linear) Hubble expansion law. It is also assumed that below the scale of statistical homogeneity, deviations from this average scenario can be described by  FLRW perturbation theory. Since there is no smooth transition between the large-scale  FLRW Hubble flow and the phenomenological Hubble flow, this latter assumption rests on quite delicate ground. For instance, the field of peculiar velocities $\{\dot{\gamma}_s(\tau)\}$ of the phenomenological observers $\left\{\tau\longrightarrow\gamma_s(\tau)\right\}$ shows a significant statistical variance \cite{Wiltshire1} with respect to the average FLRW Hubble flow and the standard of rest provided by the cosmic microwave background (CMB). 
This remark has an important effect on the relation between the celestial sphere $\mathbb{C\,S}_r(p)$ of the phenomenological observer $(p,\,\dot{\gamma}(0))$ and the corresponding celestial sphere $\widehat{\mathbb{C\,S}}_{\widehat{r}}(p)$ of the idealized FLRW observer $(p,\,\widehat{\dot{\gamma}}(0))$. They cannot be identified and must be connected by a Lorentz boost that takes into account the origin of this statistical variance.  The actual scenario is significantly constrained by the coupling of the matter inhomogeneities with a spacetime geometry that is no longer Friedmannian. As a consequence, the peculiar velocity field of the phenomenological observer may have a rather complex origin, and its variance with respect to the FLRW average expansion may become a variable of relevance in cosmography.   
This scenario naturally calls into play a delicate comparison between the geometry of  $\mathcal{C}^-(p, g)$ and the geometry of the associated FLRW past light cone that sets in at scales $L\,>\,100\,h^{-1}\,\mathrm{Mpc}$. For this purpose, along with the physical metric $g$, we  consider
 on the spacetime manifold $M$ a reference FLRW metric $\hat{g}$ and the associated family of global Friedmannian observers $\hat{\tau}\longmapsto\hat{\gamma}_s(\hat{\tau})$. Strictly speaking, the FLRW model $(M, \hat{g}, \hat{\gamma}_s(\hat{\tau}))$ should be used only over the scales $L\,>\,L_0\simeq \,100\,h^{-1}\,\mathrm{Mpc} $. We need to consider it also over the inhomogeneity scales $L\,<\,L_0$ where it plays the role of the geometrical  background  used to interpret the data according to the standard perturbative FLRW point of view recalled above. 
In such an extended role, the chosen FLRW is the \textit{Global Background Solution (GBS} according to \cite{KolbMarraMatarrese}) we need to check against the physical metric $g$ representing the phenomenological background solution. In this section, we set up the kinematical aspects for such a comparison. First some standard verbiage for introducing the FLRW model $(M, \hat{g}, \hat{\gamma}_s(\hat{\tau}))$. In terms of  the radial, and angular FRLW coordinates $y^\alpha\,:=\,\left(\hat{r}, \hat{\theta},\hat{\varphi}\right)$, and of the proper time of the comoving fundamental observers $y^4\,:=\,\hat{\tau}$, we set
\begin{eqnarray}
\widehat{g}\,&:=&\,-d\hat\tau^2\,+\,a^2(\hat\tau)\,\left[d\hat{r}^2\,+\,f^2(\hat{r})\,\left(d\hat\theta^2\,+\,\sin^2\hat\theta\,d\hat\varphi^2  \right)\right]\,,
\;\;\;\;\; \widehat{\dot\gamma}^h\,=\,\delta^h_4,\nonumber\\
\label{FLRWg}\\
 f(\hat{r})\, &:=&\,
    \begin{cases}
       & \sin\,\hat{r},\;\;\;k=+1\\
       & \hat{r},\;\;\;\;\;\;\;\;\;k=0\\
       & \sinh\,\hat{r},\;\;k=-1\,,
    \end{cases}       
\nonumber
\end{eqnarray}
where $a(\hat\tau)$ is the time-dependent scale factor, $k$ is the normalized dimensionless spatial curvature constant, and 
$\widehat{\dot\gamma}^h$ are the components of the $4$-velocity $\widehat{\dot\gamma}$ of the fundamental FLRW observers.  According to the above remarks, the  geodesics ${\tau}\longmapsto{\gamma}({\tau})$,  and $\hat{\tau}\longmapsto\hat{\gamma}(\hat{\tau})$,\,$-\delta<\, \tau,\,\hat{\tau}<\delta$, associated with the corresponding Hubble flow in $(M, g, \gamma)$ and  $(M, \hat{g}, \hat{\gamma})$, are assumed to be distinct, but in line with the scale-dependent cosmographic approach adopted here we assume that they share a common observational event $p\,\in\,M$. We denote by $\widehat{\mathscr{C}}^-(p, \hat{g})$ the associated FLRW  past light cone, and normalize the proper times $\tau$ and $\hat{\tau}$ along ${\gamma}({\tau})$ and $\hat{\gamma}(\hat{\tau})$  so that at $\tau\,=\,0\,=\,\hat{\tau}$ we have $\gamma(0)\,=\,p\,=\,\hat{\gamma}(0)$. As stressed, the two instantaneous observers $(p, \dot{\gamma}(0))$ and $(p, \widehat{\dot\gamma}(0))$ have different 4-velocities, $\dot{\gamma}(0)\not=\widehat{\dot\gamma}(0)$, and their respective  celestial spheres, $\mathbb{C\,S}(p)$ and $\widehat{C\,\mathbb{S}}^2(p)$ are quite distinct. They are related by a Lorentz trasformation describing the aberration of the sky mapping of one instantaneous observer with respect to the other. This mapping will play a basic role in our analysis, and to provide an explicit description of its properties, we start by adapting to the FLRW instantaneous observer $(p, \widehat{\dot\gamma}(0))\,\in\,(M, \hat{g}, \hat{\gamma})$ the setup characterizing the celestial spheres $\mathbb{C\,S}(p)$ and $\mathbb{C\,S}_r(p)$ of the instantaneous observer $(p, \dot{\gamma}(0))\,\in\,(M, g, \gamma)$. \\
\subsection{The FLRW celestial sphere and the associated sky sections} 
Let $\left(\widehat{T}_p M,\,\widehat{g}_p,\,\{\widehat{E}_{(i)}\}\right)$ be the tangent space to $(M, \hat{g}, \hat{\gamma})$ at $p$ endowed with a $\widehat{g}$-orthonormal frame $\{\widehat{E}_{(i)}\}_{i=1,\ldots,4}$,\,\,$\widehat{g}_p\left(\widehat{E}_{(i)}, \widehat{E}_{(k)}\right)=\eta_{ik}$, where $\eta_{ik}$ is the Minkowski metric, and where we identify $\widehat{E}_{(4)}$ with the FLRW-observer's $4$-velocity $\widehat{\dot{\gamma}}(\tau)|_{\tau=0}$,\;\,\emph{i.e.}\,\,  $\widehat{E}_{(4)}\,:=\,\widehat{\dot{\gamma}}(\tau)|_{\tau=0}$. For ease of notation, we shall often use the shorthand $\widehat{T}_p M$ when referring to the tangent space to $(M, \hat{g}, \hat{\gamma})$ at $p$. Let
\begin{equation}
\label{MinkLightConeFLRW}
C^-\left(\widehat{T}_pM,\,\{\widehat{E}_{(i)}\} \right)\,:=\,\left\{Y\,=\,\mathbb{Y}^i\widehat{E}_{(i)}\,\not=\,0\,\in\,\widehat{T}_pM\,\,|\,\widehat{g}_p(Y, Y)\,=\,0,\,\,\mathbb{Y}^4+\widehat{r}=0 \right\}\;,
\end{equation}
\begin{equation}
\label{pastMinkLightConeFLRW}
\overline{C^-\left(\widehat{T}_pM,\,\{\widehat{E}_{(i)}\} \right)}\,:=\,\left\{Y\,=\,\mathbb{Y}^i\widehat{E}_{(i)}\,\not=\,0\,\in\,\widehat{T}_pM\,\,|\,
\widehat{g}_p(Y, Y)\,\leq\,0,\,\,\mathbb{Y}^4+\widehat{r}\,\leq\,0 \right\}\;,
\end{equation}
respectively denote the set of past-directed null vectors and the set of  past-directed causal vectors in $(\widehat{T}_pM,\,\{\widehat{E}_{(i)}\})$, where $\widehat{r}:=(\sum_{a=1}^3(\mathbb{Y}^a)^2)^{1/2}$ is the radial coordinate (see (\ref{radial})) in  the hyperplane 
$\mathbb{Y}^4\,=\,0\,\subset\,\widehat{T}_pM$ parametrizing the one-parameter family of $2$-spheres 
\begin{equation}
\label{celestialRFLRW}
 \mathbb{S}^2_{\widehat{r}}(\widehat{T}_pM)\,:=\,\{Y\in C^-\left(\widehat{T}_pM, \{\widehat{E}_{(i)}\} \right)\,|\,\, \mathbb{Y}^4\,=\,-\,\widehat{r},\,\,\,\sum_{a=1}^3(\mathbb{Y}^a)^2=\widehat{r}^2,\,\,\widehat{r}\in\,\mathbb{R}_{> 0} \}\,,
 \end{equation}
that foliate $C^-\left(\widehat{T}_pM, \{\widehat{E}_{(i)}\}\right)/\{p\}$. The $2$-spheres $\mathbb{S}^2_{\widehat{r}}(\widehat{T}_pM)$, endowed with the round metric
\begin{equation}
\label{roundmetFLRW}
\widehat{\widetilde{h}}(\mathbb{S}^2)\,=\,\widehat{\widetilde{h}}_{\alpha\beta}(\mathbb{S}^2)dy^\alpha dy^\beta\,=\,d\widehat\theta^2\,+\,\sin^2\widehat\theta\,d\widehat\phi^2
\,,\,\,\,\,
0\leq\widehat\theta\leq\pi,\,\,0\leq\widehat\phi<2\pi\,
\end{equation} 
can be thought of as providing a representation of the sky, at a given value of the radial coordinate $\widehat{r}$, in the instantaneous rest space $\left(\widehat{T}_pM, \{\widehat{E}_{(i)}\} \right)$ of the FLRW  observer. In analogy with the characterization (\ref{celestialS}) of the celestial sphere $\mathbb{C\,S}(p)$, we use the projection of 
$\left.\mathbb{S}^2_{\widehat{r}}(\widehat{T}_pM)\right|_{\widehat{r}=1}$  on the hyperplane 
$\mathbb{Y}^4\,=\,0$ in $\widehat{T}_pM$, to define the FLRW \emph{celestial sphere}  
\begin{equation}
\label{celestialSFLRW}
\widehat{\mathbb{C\,S}}(p)\,\left(\left.\mathbb{S}^2_{\widehat{r}}(\widehat{T}_pM)\right|_{\widehat{r}=1},\, \widehat{\widetilde{h}}(\mathbb{S}^2)(p)\right)\,:=\,\left\{Y\,=\,\mathbb{Y}^iE_{(i)}\,\not=\,0\,\in\,\widehat{T}_pM\,\,|\,\,\mathbb{Y}^4=0,\,\,\sum_{a=1}^3(\mathbb{Y}^a)^2=1 \right\}\;,
\end{equation}
parametrizing the directions of sight 
\begin{equation}
\widehat{n}(\widehat\theta, \widehat\phi)\,:=\,(\cos\widehat\phi\sin\widehat\theta,\,\sin\widehat\phi\sin\widehat\theta,\,cos\,\widehat\theta)\,,\,\,\,\,
0\leq\widehat\theta\leq\pi,\,\,0\leq\widehat\phi<2\pi\,
\end{equation}
in the instantaneous rest space $\left(\widehat{T}_pM, \{\widehat{E}_{(i)}\} \right)$ of the FLRW observer. In full analogy with (\ref{celstSphereR}), we define   
the FLRW \emph{celestial sphere at radius $\widehat{r}$} in $\left(\widehat{T}_pM, \{\widehat{E}_{(i)}\} \right)$ according to 
\begin{equation}
\label{FLRWcelstSphereR}
\widehat{\mathbb{C\,S}}_{\widehat{r}}(p)\,:=\,\left(\mathbb{S}^2_{\widehat{r}}\left(\widehat{T}_pM\right),\,\widehat{r}^2\widehat{\widetilde{h}}(\mathbb{S}^2(p)) \right)\,.
\end{equation}
With a straightforward adaptation to the FLRW geometry of the definitions (\ref{elle}), (\ref{SachsBasis}), and (\ref{screenM}), we also introduce in $\widehat{T}_pM$ the tetrad  
\begin{equation}
\label{SachsBasisFLRW}
\left(\widehat{n}, \widehat{m}_{(2)}, \widehat{m}_{(3)}, \widehat{\ell}(\widehat{n}) \right)
\end{equation}
and associate with the pair $\left(\widehat{T}_{(\widehat{\theta}, \widehat{\phi})}\mathbb{S}^2\left(\widehat{T}_pM\right),\;\widehat{m}_{(\alpha)}(\widehat{\theta}, \widehat{\phi})\right)$  
the \emph{screen plane} $T_{\widehat{n}}\widehat{\mathbb{C\,S}}(p)$ associated with the direction of sight $\widehat{n}(\widehat{\theta}, \widehat{\phi})$ in the FLRW celestial sphere $\widehat{\mathbb{C\,S}}(p)$,
\begin{equation}
\label{FLRWscreenM}
T_{\widehat{n}}\widehat{\mathbb{C\,S}}(p)\,:=\,\left(T_{(\widehat{\theta}, \widehat{\phi})}\mathbb{S}^2\left(\widehat{T}_pM\right),\;\widehat{m}_{(\alpha)}(\widehat{\theta}, \widehat{\phi})\right)\,.
\end{equation}
Together with  the observational normal coordinates $\{X^i \}$  in $(M, g, \gamma)$, describing the local geometry on the past lightcone   $\mathscr{C}^-(p,g)\,\cap\,U_{p}$, we  introduce corresponding (normal) coordinates $\{Y^k \}$ on the past light cone $\widehat{\mathscr{C}}^-(p, \hat{g})$  in the reference FLRW spacetime  $(M, \hat{g}, \hat{\gamma})$. To begin with, let $\widehat{\exp}_p$ denote the exponential mapping based at the event $p=\hat{\gamma}(0)$, \emph{i.e.} 
\begin{eqnarray}
\label{FLRWexpmapdef}
\widehat{\exp}_p\,:\,\widehat{W}_p\,\subseteq \,\widehat{T}_pM\,&\longrightarrow&\;\;\;\;(M, \hat{g}),\\
\mathbb{Y}\;\;\;\;\;\;&\longmapsto&\;\;\;\;exp_p\,(\mathbb{Y})\,:=\,\lambda_{\mathbb{Y}}(1)\,,\nonumber
\end{eqnarray}
where $\widehat{W}_p$ is the maximal domain of $\widehat{\exp}_p$. To keep on with the notation set by (\ref{pastcone2}) and (\ref{pastcone3}),   we characterize  the  past lightcone $\widehat{\mathscr{C}}^-(p, \hat{g})\,\in\,(M, \widehat{g})$, with vertex  at $p$,\;according to
\begin{equation}
\label{pastcone2FLRW}
\widehat{\mathscr{C}}^-(p, \hat{g})\,:=\,\widehat{\exp}_p\left[\widehat{W}_p\cap C^-\left(\widehat{T}_pM, \widehat{g}_p \right)\right]\,,
\end{equation} 
and we denote by 
\begin{equation}
\widehat{\mathscr{C}}^-(p,\widehat{g};\,\widehat{r}_0)\,:=\,\left\{\left. q\,\in\,M\,\right|\,\,q\,=\,\widehat{\exp}_p(\widehat{r} \widehat{\ell}(\widehat{n}(\widehat\theta, \widehat\phi))),\, \, 0\leq \widehat{r} < \widehat{r}_0,\,\, (\widehat\theta, \widehat\phi)\in\widehat{\mathbb{C\,S}}(p)  \right\}\nonumber\,,
\end{equation}
the portion of $\widehat{\mathscr{C}}^-(p, \hat{g})$ accessible to observations for a given value $\widehat{r}_0$ of the radial parameter $\widehat{r}$.
That said, if  $\hat{U}_{p}\subset\,(M, \hat{g})$ denotes  the region of injectivity of  $\widehat{\exp}_{p}$, then normal coordinates are defined by
\begin{equation}
\label{geodnormalY}
y^i\,:=\,\mathbb{Y}^i\,\circ \,\widehat{\exp}_p^{-1}\,:\,(M, \widehat{g})\cap\,\widehat{U}_{p}\,\longrightarrow\,\mathbb{R}\,,
\end{equation}
where  $\mathbb{Y}^i$ are the components of  the vectors $\mathbb{Y}\in\,\widehat{T}_pM$ with respect to a $\hat{g}$-orthonormal frame $\{\hat{E}_{(i)}\}_{i=1,\ldots,4}$ with $\hat{E}_{(4)}\,:=\,\hat{\dot{\gamma}}(0)$. We can  parametrize
 $\widehat{\mathscr{C}}^-(p,\widehat{g})\,\cap\,\widehat{U}_{p}$ in terms of the 2-dimensional FLRW sky sections 
 \begin{equation}
 \label{FLRWsectionsky}
\widehat{\Sigma}(p, \hat{r})\,:=\,
\widehat{\exp}_p\left[\widehat{\mathbb{C\,S}}_{\widehat{r}}(p) \right]\,=\, \left\{\left.\widehat{\exp}_p\left(\widehat{r}\,\widehat{\ell}(\widehat{n}(\widehat\theta, \widehat\phi))\right)\,\right|\,\, (\widehat\theta, \widehat\phi) \,\in\,\widehat{\mathbb{C\,S}}(p)\right\}\;,
\end{equation}
 endowed with the metric induced by the inclusion of $\widehat{\Sigma}(p, \hat{r})$ into $\widehat{\mathscr{C}}^-(p, \hat{g})$, \emph{i. e.}
 \begin{equation}
 \label{etametric01}
 \left.\widehat{g}\right|_{\widehat{\Sigma}(p, \hat{r})}\,:=\,\left. (\widehat{g})_{\alpha\beta}\,dy^\alpha dy^\beta\right|_{\hat{r}}\,=\,
a^2(\widehat{\tau}(\widehat{r}))\,f^2\left(\widehat{r}\right)\left(d\widehat{\theta}^2\,+\,\sin^2\widehat\theta d\widehat{\phi}^2\right)\;,
\end{equation} 
where $a(\widehat{\tau}(\widehat{r}))$ is the  FLRW expansion factor $a(\widehat{\tau})$ (see (\ref{FLRWg})) evaluated in correspondence of the given value of the radial coordinate  $\widehat{r}\in\widehat{T}_pM$.  We proceed as in Subsection \ref{subsect2}, and exploit the exponential map $\widehat\exp_p$ to pull back  $ \left.\widehat{g}\right|_{\widehat{\Sigma}(p, \hat{r})}$ on the celestial sphere $\widehat{\mathbb{C\,S}}_{\widehat{r}}(p)$,
\begin{equation}
\label{pullbackFLRW}
\widehat{h}(\widehat{r}, \widehat\theta, \widehat\phi)\,:=\,  
\left.\left(\widehat{\exp}_p^*\,\widehat{g}|_{\widehat{\Sigma}(p, \widehat{r})}\right)_{\alpha\beta}\,dy^\alpha dy^\beta\right|_{\widehat{r}},\,\,\,\,\alpha,\,\beta\,=\,2, 3,\,\,\,\,\,\,y^2:=\widehat\theta,\,y^3:=\widehat\phi\,.
\end{equation}
This pull-back can be explicitly computed. To wit, let $y^i_q=(\widehat{r}_q, \widehat{\theta}_q, \widehat{\phi}_q, \widehat{\tau}_q)$ the normal coordinates of the event $q\in\,\widehat{\mathscr{C}}^-(p, \hat{g})$ associated with the observation of a given astrophysical source. The equation for the radial, past-directed, null geodesic connecting $q$ to the observation event $p$ reduces in the FLRW case to \cite{EllisElst}
\begin{equation}
d\widehat{r}\,=\,-\,\frac{d\widehat\tau}{a(\widehat\tau)}\,,\,\,\,\,\widehat\tau(p)\,=\,0\,=\,
\widehat{r}(p)\,,
\end{equation}
that integrates to the expression providing the (matter-comoving) radial coordinate distance between $p$ and $q$
\begin{equation}
\widehat{r}_q\,=\,\int_0^{\widehat{\tau}_q}\,\frac{d\widehat\tau}{a(\widehat\tau)}\,.
\end{equation}
Thus, the metric (\ref{pullbackFLRW}), evaluated at $\widehat{\exp_p}^{-1}(q)$, can be written in terms of $\widehat\tau_q$ as 
\begin{equation}
\widehat{h}_q\,:=\,\widehat{h}(\widehat{r}_q, \widehat\theta_q, \widehat\phi_q)\,=\, 
a^2(\widehat{\tau}_q)\,f^2\left(\widehat{r}_q\right)\left(d\widehat{\theta}_q^2\,+\,
\sin^2\widehat\theta_q d\widehat{\phi}_q^2\right)\;, 
\end{equation}
If we introduce the  dimensionless FLRW cosmological redshift corresponding to the event $q$,
\begin{equation}
\label{zetaFLRW}
z_q\,:=\,z\left(\widehat{\tau}_q\right)\,=\,\frac{a_0}{a(\widehat{\tau}_q)}\,-\,1\,,
\end{equation}
where $a_0\,:=\,a(\widehat{\tau}=0)$, then we can rewrite $\widehat{h}(\widehat{r}_q, \widehat\theta_q, \widehat\phi_q)$ as
\begin{equation}
\label{metrichatzeta}
\widehat{h}_q\,=\,
\frac{a^2_0}{(1\,+\,z_q)^2}\,f^2\left(\widehat{r}_q\right)\left(d\widehat{\theta}_q^2\,+\,
\sin^2\widehat\theta_q d\widehat{\phi}_q^2\right)\;. 
\end{equation}  
Note that the area element associated with the metric $\widehat{h}_q$,
\begin{equation}
\label{Farea}
d\mu_{\widehat{h}_q}\,=\,\frac{a^2_0}{(1\,+\,z_q)^2}\,f^2\left(\widehat{r}_q\right)\,d\mu_{\mathbb{S}^2}\,
\end{equation}
characterizes  the FLRW \emph{observer area distance} (see (\ref{Di}))  of the event $q\in\,\widehat{\mathscr{C}}^-(p, \hat{g})$ according to  
\begin{equation}
\label{FLRWDi}
\widehat{D}(\widehat{r}_q)\,=\,
\frac{a_0}{1\,+\,z_q}\,f\left(\widehat{r}_q\right)\,.
\end{equation}

\section{Comparing the celestial spheres $\mathbb{C\,S}(p)$ and $\widehat{\mathbb{C\,S}}(p)$}
 As stressed in the previous Section, the celestial sphere $\mathbb{C\,S}(p)$ of the phenomenological observer $(p, \dot\gamma(0))$, and the celestial sphere $\widehat{\mathbb{C\,S}}(p)$ of the FLRW ideal observer $(p, \widehat{\dot\gamma}(0))$ cannot be directly identified as they stand. The velocity fields $\dot{\gamma}(0)$ and $\widehat{\dot{\gamma}}(0)$ are distinct and to compensate for the induced  aberration, the celestial spheres $\mathbb{C\,S}(p)$ and $\widehat{\mathbb{C\,S}}(p)$ can be identified only up to Lorentz boosts. In the standard FLRW view, this is the familiar global  boost taking care of the kinematical dipole component in the CMB spectrum due to our peculiar motion with respect to the standard of rest provided by the CMB. However, in a cosmographic setting and presence of a complex pattern of local inhomogeneities coupled with a non-FLRW spacetime geometry over scales $\lesssim  \, 100h^{-1}\; \mathrm{Mpc}$, the peculiar motion of the phenomenological observer has a dynamical origin, driven by the gravitational interaction and not just by a kinematical velocity effect. Even if we factor out the effect of coherent bulk flows due to the non-linear local gravitational dynamics, and average the rate of expansion over spherical shells at increasing distances from $(p,\,\dot{\gamma}(0))$, the variance in the peculiar velocity of $(p,\,\dot{\gamma}(0))$  with respect to the average rate of expansion is significant \cite{Wiltshire}. These remarks imply that the Lorentz boosts connecting $\mathbb{C\,S}(p)$ and $\widehat{\mathbb{C\,S}}(p)$ acquire a dynamical meaning that plays a basic role in what follows. As a  first step, we describe the Lorentz boost in the idealized pure kinematical situation where we need to compensate for a well-defined velocity field of the celestial sphere $\mathbb{C\,S}(p)$ with respect to the celestial sphere $\widehat{\mathbb{C\,S}}(p)$ taken as providing a well-defined standard of rest. As a second step, we move to the more general setting required in the pre-homogeneity region where we sample scales $\lesssim  \, 100h^{-1}\; \mathrm{Mpc}$. In this latter case, a pure kinematical Lorentz boost will not suffice, the large fluctuations in the sources distribution require a suitable localization of the Lorentz boosts to compare the data on $\mathbb{C\,S}(p)$ with those on $\widehat{\mathbb{C\,S}}(p)$.

\subsection{The kinematical setting} 
To describe a kinematical Lorentz boost acting between $\widehat{\mathbb{C\,S}}(p)$ and $\mathbb{C\,S}(p)$,  
 we find it convenient to use in this section the well-known correspondence between the restricted Lorentz group and the six-dimensional projective special linear group $\mathrm{PSL}(2, \mathbb{C})$ describing the automorphisms of the Riemann sphere $\mathbb{S}^2\,\simeq\,\mathbb{C}\,\cup\,\{\infty\}$. More expressively, $\mathrm{PSL}(2, \mathbb{C})$ can be viewed as the group of the conformal transformations of the celestial spheres that correspond to the restricted Lorentz transformations  connecting $\mathbb{C\,S}(p)$ to $\widehat{\mathbb{C\,S}}(p)$.   In oder to set notation, let us recall that the elements of  $\mathrm{PSL}(2, \mathbb{C})$ can be identified with the set of the M\"obius transformations of the Riemann sphere $\mathbb{S}^2\,\simeq\,\mathbb{C}\,\cup\,\{\infty\}$, \emph{i.e.} the fractional linear transformations of the form
\begin{eqnarray}
\zeta\,:\,\mathbb{C}\,\cup\,\{\infty\}\,&\longrightarrow&\,\mathbb{C}\,\cup\,\{\infty\}\\
w\,&\longmapsto&\,\zeta (w)\,:=\,\frac{aw+b}{cw+d}\,,\,\,\,\,\,a, b, c, d\,\in\,\mathbb{C}\,,\,\,\,ad\,-\,bc\,\not=\,0\,,\nonumber
\end{eqnarray}
where, to avoid a notational conflict with the redshift parameter $z$, we have labeled the complex coordinate in $\mathbb{C}\,\cup\,\{\infty\}$ with  $w$ rather than with the standard  $z$.  
Let  $Y\,=\,\widehat{n}(\widehat\theta, \widehat\phi)$ denote a point on the celestial sphere $\widehat{\mathbb{C\,S}}(p)$, and let $\widehat{w}$
denote its stereographic projection\footnote{From the north pole $\theta\,=\,0\in\,\widehat{\mathbb{C\,S}}(p)$.} on the Riemann sphere $\mathbb{C}\cup\,\{\infty\}$, \emph{i.e.},
\begin{eqnarray}
\label{zetaypsilon}
&&\mathcal{P}_{\mathbb{S}^2}\,:\,\widehat{\mathbb{C\,S}}(p)\,
\longrightarrow\,\mathbb{C}\cup\{\infty \}\\
&&\mathbb{Y}^\alpha\,\longmapsto\,
\mathcal{P}_{\mathbb{S}^2}(\mathbb{Y}^\alpha)\,=\,\widehat{w}\,:=\,\frac{\mathbb{Y}^1+i\,\mathbb{Y}^2}{1-\mathbb{Y}^3}\,=\,\frac{\cos\widehat\phi\sin\widehat\theta\,+\,i\,\sin\widehat\phi\sin\widehat\theta}{1\,-\,\cos\widehat\theta}\,,\nonumber\
\end{eqnarray}
with\, $0<\theta\leq\pi,\,\,0\leq\phi<2\pi$. It is  worthwhile to stress once more that the celestial spheres $\widehat{\mathbb{C\,S}}(p)$ and ${\mathbb{C\,S}}(p)$ play the role of a mapping frame, a celestial globe where astrophysical positions are registered, and where the Lorentz boost $\widehat{\mathbb{C\,S}}(p)\,\longrightarrow\,{\mathbb{C\,S}}(p)$ must be interpreted actively as affecting only the recorded astrophysical data. In other words, the Lorentz boost affects the null directions 
in $\widehat{\mathbb{C\,S}}(p)$, mapping them in the corresponding directions in ${\mathbb{C\,S}}(p)$. To quote a few illustrative examples \cite{RindPen} of the 
$\mathrm{PSL}(2, \mathbb{C})$ transformations associated to the Lorentz group action between the celestial spheres $\widehat{\mathbb{C\,S}}(p)$ and ${\mathbb{C\,S}}(p)$, let  $v$ denote the modulus of the relative $3$-velocity of the FLRW ideal observer $(p,\,\widehat{\dot\gamma}(0))$ with respect to the
phenomenological observer $(p,\,\dot\gamma(0))$, (where ${E}^4$ is identified with the observer's 4-velocity ${\dot\gamma}(0)$). If the map between $\widehat{\mathbb{C\,S}}(p)$ and ${\mathbb{C\,S}}(p)$ is a pure Lorentz boost in a common direction, say $E^3$,  then the associated $\mathrm{PSL}(2, \mathbb{C})$ transformation is provided by 
\begin{eqnarray}
\mathrm{PSL}(2, \mathbb{C})\times \widehat{\mathbb{C\,S}}(p)\,&\longrightarrow&\,{\mathbb{C\,S}}(p)\\
\left(\zeta_{\,boost},\,\widehat{w}\right)\,&\longmapsto&\,
\zeta(\widehat{w})\,=\,w\,=\,\sqrt{\frac{1\,+\,v}{1\,-\,v}}\,\widehat{w}\,,\nonumber
\end{eqnarray}
where $\sqrt{\frac{1\,+\,v}{1\,-\,v}}$ is the relativistic Doppler factor and $w$ is the point in the Riemann sphere corresponding, under stereographic projection, to the direction $n(\theta, \phi)\in {\mathbb{C\,S}}(p)$. Similarly, 
if $\widehat{\mathbb{C\,S}}(p)$ and ${\mathbb{C\,S}}(p)$ differ by a pure rotation through an angle $\alpha$ about the $E^3$ direction, then the associated $\mathrm{PSL}(2, \mathbb{C})$ transformation is given by
\begin{eqnarray}
\mathrm{PSL}(2, \mathbb{C})\times \widehat{\mathbb{C\,S}}(p)\,&\longrightarrow&\,{\mathbb{C\,S}}(p)\\
\left(\zeta_{\,rot},\,\widehat{w}\right)\,&\longmapsto&\,
\zeta(\widehat{w})\,=\,w\,=\,e^{i\,\alpha}\,\widehat{w}\,.
\end{eqnarray}
By composing them, \emph{e. g.} by considering a rotation through an angle $\alpha$ about the $E^3$ direction, followed by a boost with rapidity $\beta\,:=\,\log\,\sqrt{\frac{1\,+\,v}{1\,-\,v}}$ along the $E^3$ axis,  
we get
\begin{eqnarray}
\label{psl2caction}
\mathrm{PSL}(2, \mathbb{C})\times \widehat{\mathbb{C\,S}}(p)\,&\longrightarrow&\,\mathbb{C\,S}(p)\\
\left( \zeta,\,\widehat{w}\right)\,&\longmapsto&\,\zeta(\widehat{w})\,=
\,w\,=\,\sqrt{\frac{1\,+\,v}{1\,-\,v}}\,e^{i\,\alpha}\,\widehat{w}\,,
\nonumber
\end{eqnarray}
describing the general fractional linear transformation mapping $\widehat{\mathbb{C\,S}}(p)$ and ${\mathbb{C\,S}}(p)$.
From the physical point of view, this corresponds to the composition 
of the adjustment of the relative orientation of the spatial bases $\{{E}_{(\alpha)}\}$ with respect to  $\{\widehat{{E}}_{(\alpha)}\}$,\;$\alpha=1,2,3$,\; followed by
a Lorentz boost adjusting for the relative velocity of $(p, \dot{\gamma}(0))$ with respect to $(p, \widehat{\dot{\gamma}}(0))$. Since the spatial directions $n(\theta, \phi)\in\,{\mathbb{C\,S}}(p)$ and $\widehat{n}(\widehat\theta, \widehat\phi)\in\,\widehat{\mathbb{C\,S}}(p)$  characterize corresponding past-directed null vectors  $\ell(\theta, \phi)\,\in\,\left(T_pM, \{E_{(i)}\} \right)$ and 
$\widehat{\ell}(\widehat\theta, \widehat\phi)\,\in\,\left(\widehat{T}_pM, \{\widehat{E}_{(i)}\} \right)$ (see (\ref{elle}) and (\ref{SachsBasisFLRW})), we can associate with the spatial directions $\{{{E}}_{(\alpha)}\}$  and $\{\widehat{{E}}_{(\alpha)}\}$ the respective null directions  
\begin{eqnarray}
\label{elleBis}
\ell_{(\alpha)}\,&=&\, E_{(\alpha)}\,-\,E_{(4)}\,=\,E_{(\alpha)}\,-\,\dot{\gamma}(0)\,,\nonumber\\
\\
\widehat{\ell}_{(\alpha)}\,&=&\, \widehat{E}_{(\alpha)}\,-\,\widehat{E}_{(4)}\,=\,\widehat{E}_{(\alpha)}\,-\,\widehat{\dot{\gamma}}(0)\,.\nonumber
\end{eqnarray}

\subsection{The pre-homogeneity setting}
From the above remarks, it follows that the Lorentz mapping  from $\widehat{\mathbb{C\,S}}(p)$ to ${\mathbb{C\,S}}(p)$ is fully determined if we specify the three distinct null directions on the FLRW celestial sphere $\widehat{\mathbb{C\,S}}(p)$ that are the images, under the $\mathrm{PSL}(2, \mathbb{C})$-transformation, of three chosen distinct sources on ${\mathbb{C\,S}}(p)$.
The selection of these three distinct sources of choice and of the corresponding null directions on ${\mathbb{C\,S}}(p)$  will depend on the scale $L$ we are probing in our cosmological observations. This is a particularly delicate matter when looking at the pre-homogeneity scales $L\,\lesssim\,100\,h^{- 1}\,\mathrm{Mpc}$, where astrophysical sources are characterized by a complex distribution of peculiar velocities with respect to the assumed Hubble flow.  To keep track of this scale dependence, let us consider the celestial spheres ${\mathbb{C\,S}}_{r}(p)$\, and\,  $\widehat{\mathbb{C\,S}}_{\widehat{r}}(p)$ defined by (\ref{celstSphereR}) and (\ref{FLRWcelstSphereR}), respectively. For  $L\,>\,0$, let  $\widehat{r}(L)$ be the value of $\widehat{r}$ such that the FLRW sky section (\ref{FLRWsectionsky}) 
\begin{equation}
 \label{FLRWsectionL}
\widehat{\Sigma}(p, \hat{r}(L))\,:=\,\widehat{\exp}_p\left[\widehat{\mathbb{C\,S}}_{\widehat{r}(L)}(p) \right]\,=\, \left\{\left.\widehat{\exp}_p\left(\widehat{r}(L)\,\widehat{\ell}(\widehat{n}(\widehat\theta, \widehat\phi))\right)\,\right|\,\, (\widehat\theta, \widehat\phi) \,\in\,\widehat{\mathbb{C\,S}}(p)\right\}\;,
\end{equation}
probes the length scale $L$. Similarly, we let ${r}(L)$ denote the value of ${r}$ such that the physical sky section (\ref{sigmapr})   
\begin{equation}
\label{sigmapr}
\Sigma(p, r(L))\,:=\,\exp_p\left[\mathbb{C\,S}_{r(L)}(p) \right]\,=\, \left\{\left.\exp_p\left(r(L)\,\ell({n}(\theta, \phi))\right)\,\right|\,\, (\theta, \phi) \,\in\,\mathbb{C\,S}(p)\right\}\,,
\end{equation}
probes the length scale $L$. Since the FLRW area distance (\ref{FLRWDi}), 
\begin{equation}
\label{setLFLRW}
\widehat{D}(\widehat{r})\,=\,
\frac{a_0}{1\,+\,z}\,f\left(\widehat{r}\right)\,,
\end{equation}
is isotropic and  may be directly expressed in terms of $z$, we may well use the redshift parameter $z$ as the reference $L$. Given $z$, we denote by $L(z)$ the corresponding length-scale of choice. As long as $\widehat{D}(\widehat{r})$ is an increasing function, we can identify $L(z)$ with the area distance $\widehat{D}(\widehat{r})$, but in general, we leave the selection of the most appropriate $L(z)$ to the nature of the cosmographical observations one wants to perform. Given $\zeta\in\mathrm{PSL}(2, \mathbb{C})$ and a  value of the redshift $z$, we have a corresponding relation between the "radial" variables $\widehat{r}(L(z))$ and $r(L(z))$ in (\ref{FLRWsectionL}) and (\ref{sigmapr}). We can take advantage of this relation to simplify the notation for the celestial spheres and the associated sky sections according to 
\begin{equation}
 \label{FLRWsigmaL}
\widehat{\mathbb{C\,S}}_z(p)\,:=\,\widehat{\mathbb{C\,S}}_{\widehat{r}(L(z))}(p)\,\Longrightarrow\,
\widehat{\Sigma}_z\,:=\,\widehat{\Sigma}(p, \hat{r}(L(z)))\,:=\,\widehat{\exp}_p\left[\widehat{\mathbb{C\,S}_z}(p) \right]\,,
\end{equation}
and
\begin{equation}
\label{sigmaL}
\mathbb{C\,S}_z(p)\,:=\,\mathbb{C\,S}_{r(L(z))}(p)\,\Longrightarrow\,
\Sigma_z\,:=\,\Sigma(p, r(L(z)))\,:=\,\exp_p\left[\mathbb{C\,S}_z(p) \right]\,,
\end{equation}
a notation  that, if not otherwise stated, we adopt henceforth. Since in the pre-homogeneity region $L(z)\,\lesssim\,100\,h^{-1}\,\mathrm{Mpc}$, the large variance in peculiar velocities of the astrophysical sources implies a great variability in the selection of the three reference null directions that fix the $\mathrm{PSL}(2, \mathbb{C})$ action, we localize this action according to the following construction.
\begin{itemize}
\item{We assume that there is a finite collection of  points $\{y_{(I)}\}\,\in\,\widehat{\mathbb{C\,S}}_z(p)$ and a corresponding collection of open disks $\{\widehat{B}(y_{(I)}, \delta) \}$ of radius $\delta$, centered at the points $\{y_{(I)}\}$, and defined by 
\begin{equation}
\label{hatbin}
\widehat{B}(y_{(I)}, \delta)\,:=\, \{y'\in \widehat{\mathbb{C\,S}}_z(p)\,|\,d_{\mathbb{S}^2}(y',y_{(I)})\,\leq \,\delta\} \subset \widehat{\mathbb{C\,S}}_z(p)\,
\end{equation}
 where $d_{\mathbb{S}^2}(y',y_{(I)})$ denotes the distance in the round unit metric on $\mathbb{S}^2$. We also assume that any such $\widehat{B}(y_{(I)}, \delta)$ contains the images of three reference astrophysical sources of choice, call them $A_{(I,\,k)}$, \,$k=1,2,3,$,\, with celestial coordinates in $\widehat{\mathbb{C\,S}}_z(p)$ given by  $y_{(I,\,k)}\,=:\,\widehat{n}_{(I,\,k)}(\widehat{\theta}, \widehat{\phi})$.  }
\item{We adopt a similar partition on the celestial sphere $\mathbb{C\,S}_z(p)$, to the effect that associated with each disk $\widehat{B}(y_{(I)}, \delta)$ there is, in $\mathbb{C\,S}_z(p)$,  a corresponding metric disk
\begin{equation}
B(x(y_{(I)}),\,\delta)\,=\,\{x'\in \mathbb{C\,S}_z(p) \,\,|\,d_{\mathbb{S}^2}(x', x(y_{(I)}))\leq  \delta\}\,\subset\, \mathbb{C\,S}_z(p)\,.
\end{equation}
We require that the images $A_{(I,\,k)}$ of the three reference astrophysical sources of choice, that in $\widehat{B}(y_{(I)}, \delta)$ have celestial coordinates $y_{(I,\,k)}$, are represented in $B(x(y_{(I)}),\,\delta)$ by three distinct points with celestial coordinates $x_{(I,\,k)}\,=:\,{n}_{(I,\,k)}({\theta},{\phi})$.   }
\item{We further assume that the past null directions $\widehat{\ell}_{(I,\,k)}\,=\,\widehat{n}_{(I,\,k)}(\widehat{\theta}, \widehat{\phi})\,-\,\widehat{\dot\gamma}(0)$, associated with the location of the reference sources $A_{(I,\,k)}$ in the portion of the celestial sphere $\widehat{B}(y_{(I)}, \delta)\cap\widehat{\mathbb{C\,S}}_z(p)$, are related to the corresponding null directions ${\ell}_{(I,\,k)}\,=\,{n}_{(I,\,k)}({\theta}, {\phi})\,-\,{\dot\gamma}(0)$, locating the sources $A_{(I,\,k)}$ in 
${B}(x(y_{(I)}), \delta)\cap{\mathbb{C\,S}}_z(p)$, by the $\mathrm{PSL}(2, \mathbb{C})$ map
\begin{eqnarray}
\label{ZetaLoc}
\zeta_{(I)}\,:\, \widehat{B}(y_{(I)}, \delta)\cap\widehat{\mathbb{C\,S}}_z(p)\,&\longrightarrow&\,{B}(x(y_{(I)}), \delta)\cap\mathbb{C\,S}_z(p)\\
\widehat{w}\,&\longmapsto&\,\zeta_{I}(\widehat{w})\,=
\,w\,=\,\sqrt{\frac{1\,+\,v}{1\,-\,v}}\,e^{i\,\alpha(A_{(I,\,k)})}\,\widehat{w}\,,
\nonumber
\end{eqnarray} 
where  $\sqrt{\frac{1\,+\,v}{1\,-\,v}}\,e^{i\,\alpha(A_{(I,\,k)})}$ is the composition of the Lorentz boost ($v$ being the relative 3-velocity of $\dot\gamma(0)$ with respect to $\widehat{\dot\gamma}(0)$) and of the spatial rotation that, at the given scale $L(z)$, allow us to align the portion of the celestial sphere $\mathbb{C\,S}_z(p)$ described by ${B}(x(y_{(I)}), \delta)$ with the portion of the FLRW celestial sphere  $\widehat{\mathbb{C\,S}}_z(p)$ described by $\widehat{B}(y_{(I)}, \delta)$. }
\item{Finally, we require that the finite collections of  \emph{celestial coordinate bins} $\{\widehat{B}(y_{(I)}, \delta)\}$ and $\left\{{B}(x(y_{(I)}), \delta)\right\}$ cover the respective celestial spheres $\widehat{\mathbb{C\,S}}_z(p)$ and $\mathbb{C\,S}_z(p)$.}
\end{itemize}
It is worthwhile to stress that the collections of  bins $\{\widehat{B}(y_{(I)}, \delta)\}$ and $\left\{{B}(x(y_{(I)}), \delta)\right\}$ can be chosen in many distinct ways, according to the cosmographic observations one wishes to carry out (we use disks for mathematical convenience). Whatever choice of the above type we make, we can extend the localized $\mathrm{PSL}(2, \mathbb{C})$ maps (\ref{ZetaLoc}) by using a smooth partition of unity $\left\{\chi_{(I)}  \right\}$ subordinated to the finite covering  $\{\widehat{B}(y_{(I)}, \delta)\}$ of  $\widehat{\mathbb{C\,S}}_z(p)$, \emph{i.e.} a collection of smooth functions 
$\chi_{(I)}\,:\,\widehat{B}(y_{(I)}, \delta)\longrightarrow\,[0, 1]$ whose support is such that $\mathrm{supp}\,\chi_{(I)}\subseteq\,\widehat{B}(y_{(I)}, \delta)$ and such that 
$\sum_{y\in\,\widehat{\mathbb{C\,S}}_z(p)}\,\chi_{(I)}(y)\,=\,1$. We define the \emph{localized $\mathrm{PSL}(2,\mathbb{C})$ map} connecting, at scale $L(z)$, the celestial spheres 
$\widehat{\mathbb{C\,S}}_z(p)$ and ${\mathbb{C\,S}}_z(p)$, decorated with the respective coordinate bins $\{\widehat{B}(y_{(I)}\}$ and $\{{B}(x(y_{(I)}), \delta)\}$, according to
\begin{eqnarray}
\label{ZetaL}
\zeta_{(z)}\,:\, \widehat{\mathbb{C\,S}}_z(p)\,&\longrightarrow&\,\mathbb{C\,S}_z(p)\\
\widehat{w}\,&\longmapsto&\,\zeta_{(z)}(\widehat{w})\,:=\,
\sum_{y\in\,\widehat{\mathbb{C\,S}}_z(p)}\,\chi_{(I)}(y)\,\zeta_{(I)}(w)\,,
\nonumber
\end{eqnarray}
where $\zeta_{(I)}(w)$ is provided by (\ref{ZetaLoc}). Note that, when necessary, this localized $\mathrm{PSL}(2,\mathbb{C})$ map can be further generalized by completing it in the Sobolev space of maps which together with their derivatives are square-summable over 
$\widehat{\mathbb{C\,S}}_z(p)$. This completion requires some care which we do not enter here (see \cite{CarFam} for details), and it is needed when discussing the  distance between the FLRW and the cosmographic lightcones.\\

It is worthwhile to stress that in the pre-homogeneity region $L(z)\,\lesssim\,100\,h^{-1}\,\mathrm{Mpc}$, the large variance in peculiar velocities of the astrophysical sources implies a great variability in the selection of the three reference null directions that fix the local $\mathrm{PSL}(2, \mathbb{C})$ action characterizing the map $\zeta_{(z)}$. This implies that  $\zeta_{(z)}$ may vary considerably with $L(z)$. Recall that the role of the celestial spheres 
${\mathbb{C\,S}}_z(p)$ and $\widehat{\mathbb{C\,S}}_z(p)$ is simply that of representing past null directions at the observational event $p\,\in\,M$, directions that respectively point to the astrophysical sources on the sky section ${\Sigma}_z$, as seen by $(p, \dot{\gamma}(0))$, and on $\widehat{\Sigma}_z$, as seen according to $(p, \widehat{\dot{\gamma}}(0))$. These data are transferred from these sky sections to the respective celestial spheres through null geodesics, thus we can associate with the localized $\mathrm{PSL}(2, \mathbb{C})$ action the map between the sky sections $\widehat{\Sigma}_z$ and ${\Sigma}_z$ given by
\begin{eqnarray}
\label{psimap}
\psi_{(z)}\,:\,\widehat{\Sigma}_z\,&\longrightarrow&\,\Sigma_z\\
q\,&\longmapsto&\,\psi_{(z)}(q)\,:=\,\exp_p\,\circ\,\zeta_{(z)}\,\circ\,\widehat{\exp}_p^{\,-1}(q)\,,\nonumber
\end{eqnarray}
for any point $q\,\in\,\widehat{\Sigma}_z$.

\section{The comparison between the screen planes $T_{\widehat{n}}\widehat{C\,S}_z(p)$ and $T_{n}{C\,S}_z(p)$}
\label{HEFsect}
The localized $\mathrm{PSL}(2, \mathbb{C})$ map $\zeta_{(z)}$ induces a corresponding map between the
\emph{screen plane} $T_{\widehat{n}}\widehat{\mathbb{C\,S}}(p)_z$ associated with the direction of sight $\widehat{n}(\widehat{\theta}, \widehat{\phi})$ in the FLRW celestial sphere $\widehat{\mathbb{C\,S}}_z(p)$ (see (\ref{FLRWscreenM})), and the \emph{screen plane} $T_{n}\mathbb{C\,S}_z(p)$ associated with the direction of sight ${n}(\theta, \phi)\,=\,\zeta_{(z)}\left(\widehat{n}(\widehat{\theta}, \widehat{\phi})  \right)$ in the celestial sphere $\mathbb{C\,S}_z(p)$ (see (\ref{screenM})).
The geometry of this correspondence is quite sophisticated since it is strictly related to harmonic map theory and it will be described here in some detail. To begin with, we denote by $T\widehat{\mathbb{C\,S}}_z$ and by $T\mathbb{C\,S}_z$ the \emph{screen bundles} associated with the screen planes on $\widehat{\mathbb{C\,S}}_z(p)$  and $\mathbb{C\,S}_z(p)$, respectively. These are just two copies of the usual tangent bundle $T\mathbb{S}^2$ of the 2-sphere. If there is no danger of confusion, we use both notations in what follows. Under such notational assumptions, we can associate with the map (\ref{ZetaL}),
\begin{equation}
\label{ZetaL2}
\zeta_{(z)}\,:\, \widehat{\mathbb{C\,S}}_z(p)\,\longrightarrow\,\mathbb{C\,S}_z(p)\,,
\end{equation}
the  pull--back  bundle $\zeta_{(z)}^{-1}\,T\mathbb{C\,S}_z$  whose sections $\mathrm{v}\equiv \zeta_{(z)}^{-1}V:= V\circ \zeta_{(z)}$,\,  $V\in C^{\infty }(\mathbb{C\,S}_z(p), T\mathbb{C\,S}_z)$,\, are the vector fields over $\mathbb{C\,S}_z(p)$ covering the map $\zeta_{(z)}$. In physical terms, the vectors $v$ are the tangent vector on the celestial sphere $\mathbb{C\,S}_z(p)$ that describe the (active) effect of the combination of rotation and Lorentz boost induced by $\zeta_{(z)}$ on the null direction $\widehat\ell(\widehat{n})$. More expressively, let us remark that for a given direction of sight $\zeta_{(z)}(\widehat{n})\,=\,n(\theta,\phi)\,\in\,\mathbb{C\,S}_z(p)$, the vectors $V\in T_{n}\mathbb{C\,S}_z(p)$ can be used to describe the geometrical characteristics of the astrophysical images on the screen $T_{n}\mathbb{C\,S}_z(p)$, for instance, the apparent diameters of the source. Thus, the vectors $\mathrm{v}\equiv \zeta_{(z)}^{-1}V:= V\circ \zeta_{(z)}$, sections of the  pull--back  bundle $\zeta_{(z)}^{-1}\,T\mathbb{C\,S}_z$, can be interpreted as transferring the "images" of the screens in $T\mathbb{C\,S}_z$ back to $\widehat{\mathbb{C\,S}}_z(p)$ so as to be able to compare them with the reference screen-shots in $T\widehat{\mathbb{C\,S}}_z$.    In terms of the local coordinates $y^{a}\,:=\,\left(\widehat\theta, \widehat\phi  \right)$,\;$a=1,2$,\; on $\widehat{\mathbb{C\,S}}_z(p)$ (see (\ref{geodnormalY}))\footnote{In what follows the $(\widehat\theta, \widehat\phi)$, corresponding to $(y^2, y^2)$ in the normal coordinates string  $\{y^\alpha\}$, are relabelled as $\{y^a\}$, with $a=1,2$; a similar relabeling is also adopted for the normal coordinates $(\theta, \phi)$ on ${\mathbb{C\,S}}_z(p)$. },   we can write the section $\mathrm{v}\equiv \zeta_{(z)}^{-1}V:= V\circ \zeta_{(z)}$ as\footnote{In what follows we freely refer to the excellent \cite{Eells}, \cite{helein}, and \cite{jost} for a detailed analysis of the geometry of the computations involved in harmonic map theory.}
\begin{equation}
{\mathbb{C\,S}}_z(p)\,\ni \,y^a\,\longmapsto \,\mathrm{v}(y^a)\,=\,\mathrm{v}^b(y)\,\frac{\partial}{\partial \zeta_{(z)}^b(y)}\,\in \,\left.\zeta_{(z)}^{-1}T\mathbb{C\,S}_z\right|_y\;,
\end{equation}
where $\zeta_{(z)}^b(y)$, \,$b\,=1,2$, are the coordinates of the point (direction of sight) in $\zeta_{(z)}(y)\,\in{\mathbb{C\,S}}_z(p)$  given, in terms of the $y^a$ by (\ref{zetaypsilon}).
In particular, if $T^*\widehat{\mathbb{C\,S}}_z$ denotes the cotangent bundle to $\widehat{\mathbb{C\,S}}_z(p)$, we can locally introduce the differential
\begin{equation}
\label{zetaDifferential}
d\zeta_{(z)}\,=\,\frac{\partial\zeta_{(z)}^b}{\partial y^a}dy^a\otimes \frac{\partial }{\partial \zeta_{(z)}^b}\;,
\end{equation}
and interpret it as a section of the product bundle  $T^*\widehat[\mathbb{C\,S}]_z\otimes \zeta_{(z)}^{-1}\,T\mathbb{C\,S}_z$. To provide a comparison between the geometrical information gathered from the astrophysical data, let us recall that on the screens $T\widehat{\mathbb{C\,S}}_z$ and $T\mathbb{C\,S}_z$ we have the inner products respectively defined by the pull-back metrics (\ref{pullbackFLRW}) and (\ref{metrich}), \emph{i.e.}
\begin{equation}
\label{pullbackFLRW2}
\widehat{h}(\widehat{r}(L(z)), \widehat\theta, \widehat\phi)\,:=\,  
\left.\left(\widehat{\exp}_p^*\,\widehat{g}|_{\widehat{\Sigma}_z}\right)_{ab}\,dy^a dy^b\right|_{\widehat{r}(L(z))},\,\,\,\,a,\,b\,=\,1, 2,\,\,\,\,\,\,y^1:=\widehat\theta,\,y^2:=\widehat\phi\,.
\end{equation}
and
\begin{equation}
\label{metrich2}
h(r(L(z)),\theta, \phi)\,:=\,\left.\left(\exp_p^*\,g|_{\Sigma_z}\right)_{ab}\,dx^a dx^b\right|_{r(L(z))},\,\,\,\,a,\,b\,=\,1, 2,\,\,\,\,\,\,x^1:=\theta,\,x^2:=\phi\,.
\end{equation}
The Riemannian metric in the pull-back screen  $ \left(\zeta_{(z)}^{-1}\,T\mathbb{C\,S}_z\right)_y$ over $y\in \widehat{\mathbb{C\,S}}_z(p)$ is provided by $h(\zeta_{(z)}(y))$, hence the tensor bundle  $T^*\widehat{\mathbb{C\,S}}_z\otimes \zeta_{(z)}^{-1}\,T\mathbb{C\,S}_z$ over the celestial sphere $\widehat{\mathbb{C\,S}}_z(p)$ is endowed with the pointwise inner product  
\begin{equation}
\label{bundmetr}
\langle\cdot ,\cdot \rangle_{T^*\widehat[\mathbb{C\,S}]_z\otimes \zeta_{(z)}^{-1}\,T\mathbb{C\,S}_z}\,:=\,
\widehat{h}^{-1}(y)\otimes h(\zeta_{(z)}(y))(\cdot ,\cdot )\;,
\end{equation}
where $\widehat{h}^{-1}(y)\,:=\,\widehat{h}^{ab}(y)\,\partial_a \otimes \partial_b$ is the metric tensor in $T^*_y\widehat{\mathbb{C\,S}}_z$. 
The corresponding Levi-Civita connection will be denoted by $\nabla^{\langle,\rangle}$. Explicitly,  if $W\,=\,W_a^b\,dy^a\otimes \frac{\partial }{\partial \zeta_{(z)}^b}$ is a section of  $T^*\widehat{\mathbb{C\,S}}_z\otimes \zeta_{(z)}^{-1}\,T\mathbb{C\,S}_z$, the covariant derivative of $W$ in the direction $\frac{\partial}{\partial y^b}$ is provided by
\begin{eqnarray}
\label{indConn}
&&\nabla^{\langle,\rangle}_b\,W\,=\,\nabla^{\langle,\rangle}_b\,\left( W_a^c\,dy^a\otimes \frac{\partial }{\partial \zeta_{(z)}^c}\right)\\
&&=\,\frac{\partial}{\partial y^b}\,W_a^c\,dy^a\otimes \frac{\partial }{\partial \zeta_{(z)}^c}+W_a^c\left(\widehat\nabla_b\,dy^a \right)\otimes \frac{\partial }{\partial \zeta_{(z)}^c} \nonumber\\
&&+\,W_a^c\,dy^a \otimes \,\nabla^{*}_b\left(\frac{\partial }{\partial \zeta_{(z)}^c}\right)\nonumber\;,
\end{eqnarray}
where $\widehat\nabla$   denotes the Levi--Civita connection on $(\widehat{\mathbb{C\,S}}_z(p),\widehat{h})$, and $\nabla^{*}$ is the pull back on  $\zeta_{(z)}^{-1}\,T\mathbb{C\,S}_z$ of  the Levi--Civita connection of $({\mathbb{C\,S}}_z, h)$. If $\widehat{\Gamma}^a_{bc}(\widehat{h})$ and $\Gamma^a_{bc}(h)$ respectively denote the Christoffel symbols of $(\widehat{\mathbb{C\,S}}_z(p),\widehat{h})$ and $({\mathbb{C\,S}}_z(p), h)$, then  $\widehat{\nabla}_b\,dy^a\,=\,-\,\widehat{\Gamma}_{bc}^a(\widehat{h})\,dy^c$ and  $\nabla^{*}_b\left(\frac{\partial }{\partial \zeta_{(z)}^c}\right)\,=\,\frac{\partial \zeta_{(z)}^i}{\partial y^b}\,\Gamma^k_{ci}(h)\,\frac{\partial }{\partial \zeta_{(z)}^k}$, and one computes
\begin{equation}
\nabla^{\langle,\rangle}_b\,W\,=\,\left(\frac{\partial}{\partial y^b}\,W_a^i\,-\,W^i_c\widehat{\Gamma}^c_{ba}(\widehat{h})\,+\,W^k_a\frac{\partial \zeta_{(z)}^j}{\partial y^b}\,\Gamma_{kj}^i(h)  \right)\,dy^a\otimes \frac{\partial }{\partial \zeta_{(z)}^i}\;.
\end{equation}
These remarks on the geometry of the map (\ref{ZetaL2}) allow us to  compare the data on the screens $T\mathbb{C\,S}_z$ and $T\widehat{\mathbb{C\,S}}_z$. For this purpose, the relevant quantity is the norm, evaluated with respect to the inner product (\ref{bundmetr}), of the differential (\ref{zetaDifferential}) of the $\mathrm{PSL}(2, \mathbb{C})$ map $\zeta_{(z)}$. Direct computation provides
\begin{eqnarray}
\label{HSnormZeta}
e(\widehat{h}, \zeta_{(z)}; h)\,&:=&\,
\langle d\zeta_{(z)} , d\zeta_{(z)} \rangle_{T^*\widehat[\mathbb{C\,S}]_z\otimes \zeta_{(z)}^{-1}\,T\mathbb{C\,S}_z}\\
&=&\,
\widehat{h}^{ab}(y)\,\frac{\partial \zeta_{(z)}^{i}(y)}{\partial y^a}
\frac{\partial\zeta_{(z)}^{j}(y)}{\partial y^b}\,h_{ij}(\zeta_{(z)}(y))\,=\,tr_{\widehat{h}(y)}\,(\zeta_{(z)}^{*}\,h)\,,\nonumber
\end{eqnarray} 
where $tr_{\widehat{h}(y)}\,(\zeta_{(z)}^{*}\,h)$ denotes the trace, with respect to the metric $\widehat{h}$ of the pull-back metric $\zeta_{(z)}^{*}\,h$. In other words, at any point $y$, $e(\widehat{h}, \zeta_{(z)}; h)(y)$ is the sum of the eigenvalues of the metric $\zeta_{(z)}^{*}\,h$, thus providing the sum of the squares of the length stretching generated 
by the (pull-back of) the physical metric $\zeta_{(z)}^{*}\,h$  along the orthogonal directions $(\widehat\theta, \widehat\phi)$. To such stretching,  we can associate the \emph{tension field} of the map $\zeta_{(z)}$, defined by
\begin{equation}
\label{tensionZeta}
\tau^i(\zeta_{(z)})\,:=\,\Delta_{(\widehat{h})}\,\zeta_{(z)}^i\,+\,
\widehat{h}^{kj}\,\Gamma_{ab}^i(h)\, \frac{\partial \zeta_{(z)}^a}{\partial y^k}\frac{\partial \zeta_{(z)}^b}{\partial y^j}\,.
\end{equation}
To provide some intuition on these geometrical quantities, we can adapt to our case a nice heuristic remark by J. Eells and  L. Lemaire described in their classical paper on harmonic map theory \cite{Eells}. Let us imagine the FLRW celestial sphere $(\widehat{\mathbb{C\,S}}_z(p),\widehat{h})$ as a rubber balloon, decorated with dots representing the astrophysical sources recorded from the sky section $\widehat\Sigma_z$. This balloon has the geometry described by the round metric $\widehat{h}(z, \widehat\theta, \widehat\phi)$ defined by (\ref{pullbackFLRW2}), explicitly (see (\ref{metrichatzeta}))
\begin{equation}
\label{metrichatzeta2}
\widehat{h}(\widehat{r}(z), \widehat\theta, \widehat\phi)\,=\,
\frac{a^2_0}{(1\,+\,z_{L})^2}\,f^2\left(\widehat{r}(z)\right)\left(d\widehat{\theta}^2\,+\,
\sin^2\widehat\theta d\widehat{\phi}^2\right)\;, 
\end{equation}  
where $z_{L}$ is the redshift associated with the length scale $L$.
Conversely, let us imagine the physical celestial sphere 
$({\mathbb{C\,S}}_z(p),\,{h})$ as a rigid surface with the geometry induced by the metric $h(r(z), \theta, \phi)$ defined by  (\ref{metrich2}), \emph{i.e.}, (see (\ref{etametric0})),
\begin{eqnarray}
\label{etametric02}
&&h\left(r(z), \theta, \phi\right)\\
&&\,=\, D^2(r(z), \theta, \phi)\left(d{\theta}^2\,+\,
\sin^2\theta d{\phi}^2\,+\,\mathcal{L}_{ab}(r(z), \theta, \phi)\,dx^a dx^b \right)\,,\;\;\;x^1:=\theta,\,x^2:=\phi\,,\nonumber
\end{eqnarray}
providing the geometric landscape of the astrophysical sources reaching us along null geodesics from the physical sky section $\Sigma_z$. We can think of  the $PSL(2, \mathbb{C})$ map $\zeta_{(z)}$ as stretching the elastic surface $(\widehat{\mathbb{C\,S}}_z(p),\widehat{h})$ on the rigid surface  $({\mathbb{C\,S}}_z(p),\,{h})$. The purpose of this stretching is to overlap the images of the astrophysical sources recorded on $(\widehat{\mathbb{C\,S}}_z(p),\widehat{h})$ with the images of the corresponding sources as registered on $({\mathbb{C\,S}}_z(p),{h})$. In general, this overlap is not successful without stretching the surface,  and to any point $y\,\in\,(\widehat{\mathbb{C\,S}}_z(p),\widehat{h})$ we can associate a corresponding vector measuring the stretch necessary for connecting the  images of the same source on the two celestial spheres\footnote{This is not to be confused with the phenomenon of strong gravitational lensing that occurs in a given celestial sphere. It is simply a mismatch due to the comparison between the description of the same astrophysical source on two distinct celestial spheres.} $(\widehat{\mathbb{C\,S}}_z(p),\widehat{h})$ and $({\mathbb{C\,S}}_z(p),{h})$. To leading order, the required stretching is provided by the tension vector $\tau^i(\zeta_{(z)}, y)$ at $y$.  
Both the Hilbert-Schmidt norm (\ref{HSnormZeta}) and the tension vector field (\ref{tensionZeta}) of the map $\zeta_{(z)}$ are basic quantities in harmonic map theory, and to understand the strategy we will follow in comparing, at a given length scale $L$, the FLRW past light cone $\widehat{\mathcal{C}}(p, \widehat{g})$ with the physical observational past light cone ${\mathcal{C}}(p, {g})$ we need to look into the harmonic map theory associated with $\zeta_{(z)}$. Let us start by associating with $\langle d\zeta_{(z)} , d\zeta_{(z)} \rangle_{T^*\widehat[\mathbb{C\,S}]_z\otimes \zeta_{(z)}^{-1}\,T\mathbb{C\,S}_z}$ the density 
\begin{equation}
\label{HEdensity}
e(\widehat{h}, \zeta_{(z)}, h)\,\,d\mu_{\widehat{h}}\,:=\,\langle d\zeta_{(z)} , d\zeta_{(z)} \rangle_{T^*\widehat[\mathbb{C\,S}]_z\otimes \zeta_{(z)}^{-1}\,T\mathbb{C\,S}_z}\, d\mu_{\widehat{h}}\,
=\,tr_{\widehat{h}(y)}\,(\zeta_{(z)}^{*}\,h)\,d\mu_{\widehat{h}}\;,
\end{equation}
where $d\mu_{\widehat{h}}$ is the volume element defined by the metric $\widehat{h}$ on the FLRW celestial sphere  $\widehat{\mathbb{C\,S}}_z(p)$.
An important property of the density $e(\widehat{h}, \zeta_{(z)}; h)\,\,d\mu_{\widehat{h}}$ is that it is invariant under the two-dimensional conformal transformations 
\begin{equation}
\label{conftrsf}
(\widehat{\mathbb{C\,S}}_z(p),\, \widehat{h}_{ab})\,\longmapsto\, (\widehat{\mathbb{C\,S}}_z(p),\, e^{- f}\,\widehat{h}_{ab})\;,
\end{equation}
where $f$ is a smooth function on $\widehat{\mathbb{C\,S}}_z(p)$. In this connection, it is worthwhile to recall that conformal invariance is strictly related to the action of the Lorentz group on the celestial spheres (and it is ultimately the rationale for the relation between Lorentz transformations and the fractional linear transformations of $\mathrm{PSL}(2, \mathbb{C})$).\\
The expression  
$\frac{1}{2}\,e(\widehat{h}, \zeta_{(z)}; h)\,\,d\mu_{\widehat{h}}$ characterizes the harmonic map energy functional associated to the map $\zeta_{(z)}$, \emph{viz.} 
\begin{equation}
E[\widehat{h}, \zeta_{(z)},\,h]\,:=\,\frac{1}{2}\,\int_{\widehat{\mathbb{C\,S}}_z}\,
e(\widehat{h}, \zeta_{(z)}, h)\,\,d\mu_{\widehat{h}}\;.
\label{Msigmamod0}
\end{equation}
It is worthwhile to put forward  a more explicit characterization of the nature of the harmonic map functional $E[\widehat{h}, \zeta_{(z)};\,h]$ by making explicit, together with 
the celestial spheres $\widehat{\mathbb{C\,S}}_z(p)$ and ${\mathbb{C\,S}}_z(p)$, the role of the corresponding sky sections $\widehat{\Sigma}_z$ and $\Sigma_z$. To this end, let us consider the map (\ref{psimap})  acting between the sky sections $\widehat{\Sigma}_z$ and ${\Sigma}_z$,  
\begin{eqnarray}
\label{psimap2}
\psi_{(z)}\,:\,(\widehat{\Sigma}_z,\, \left.\widehat{g}\right|_{\hat{\Sigma}_z})\,&\longrightarrow&\,(\Sigma_z,\,
\left.{g}\right|_{{\Sigma}_z})\\
y\,&\longmapsto&\,\psi_{(z)}(q)\,:=\,\exp_p\,\circ\,\zeta_{(z)}\,\circ\,\widehat{\exp}_p^{\,-1}(y)
\,.\nonumber
\end{eqnarray}
The corresponding harmonic map functional is provided by
\begin{equation}
E\left[\widehat{g}_{(z)},\, \psi_{(z)},\,{g}_{(z)}\right]\,:=\,\frac{1}{2}\,\int_{\widehat\Sigma_z}\,
(\widehat{g}_{(z)})^{ab}\,\frac{\partial\psi_{(z)}^i(y)}{\partial y^a}\frac{\partial\psi_{(z)}^k(y)}{\partial y^b}\,(g_{(z)})_{ik}\,d\mu_{\widehat{g}_{(z)}}
\end{equation}
where, for notational ease, we have set $\widehat{g}_{(z)}\,:=\,\widehat{g}_{\,\hat{\Sigma}_z}$ and ${g}_{(z)}\,:=\,{g}_{\,{\Sigma}_z}$. We can equivalently write $E\left[\widehat{g}_{(z)},\, \psi_{(z)},\,{g}_{(z)}\right]$ in terms of pull-backs of the relevant maps, and get the following chain of relations 
\begin{eqnarray}
\label{chainrelation}
E\left[\widehat{g}_{(z)},\, \psi_{(z)},\,{g}_{(z)}\right]\,&=&\,\frac{1}{2}\,\int_{\widehat\Sigma_z}\,
(\widehat{g}_{(z)})^{ab}\,\left(\psi_{(z)}^*g_{(z)} \right)_{ab}\,d\mu_{\widehat{g}_{(z)}}\\
&=&\,\frac{1}{2}\,\int_{\widehat{\exp_p}(\widehat{\mathbb{C\,S}}_z)}\,
(\widehat{g}_{(z)})^{ab}\,\left(\psi_{(z)}^*g_{(z)} \right)_{ab}\,d\mu_{\widehat{g}_{(z)}}\nonumber\\
&=&\,\frac{1}{2}\,\int_{\widehat{\mathbb{C\,S}}_z}\,
\widehat{\exp_p}^*\left[
(\widehat{g}_{(z)})^{ab}\,\left(\psi_{(z)}^*g_{(z)} \right)_{ab}\right]\,\widehat{\exp_p}^*(d\mu_{\widehat{g}_{(z)}})\nonumber\\
&=&\,\frac{1}{2}\,\int_{\widehat{\mathbb{C\,S}}_z}\,
\widehat{h}^{ab}\,\left(\widehat{\exp_p}^*\,\left(\psi_{(z)}^*g_{(z)}\right) \right)_{ab}\,d\mu_{\widehat{h}}\nonumber\\
&=&\,\frac{1}{2}\,\int_{\widehat{\mathbb{C\,S}}_z}\,
\widehat{h}^{ab}\,\left(\widehat{\exp_p}^*\,\left(\exp_p\,\circ\,\zeta_{(z)}\,\circ\,\widehat{\exp}_p^{\,-1}\right)^*g_{(z)} \right)_{ab}\,d\mu_{\widehat{h}}\nonumber\\
&=&\,\frac{1}{2}\,\int_{\widehat{\mathbb{C\,S}}_z}\,
\widehat{h}^{ab}\,\left(\zeta_{(z)}^*h \right)_{ab}\,d\mu_{\widehat{h}}\,=\,E[\widehat{h},\, \zeta_{(z)},\,{h}]\nonumber\,,
\end{eqnarray}
from which it follows that the harmonic map energy functional associated with the localized $\mathrm{PSL}(2, \mathbb{C})$ map $\zeta_{(z)}$  and with the map $\psi_{(z)}$, defined by (\ref{psimap2}), can be identified. This is not surprising since $\psi_{(z)}\,:=\,\exp_p\,\circ\,\zeta_{(z)}\,\circ\,\widehat{\exp}_p^{\,-1}$
can be seen as the representation of $\zeta_{(z)}$ on the sky sections $\widehat{\Sigma}_z:=\widehat{\exp}_p\left(\widehat{\mathbb{C\,S}}_z(p)\right)$ and ${\Sigma}_z:={\exp}_p\left({\mathbb{C\,S}}_z(p)\right)$. 
From the conformal nature of the map  $\zeta_{(z)}:\widehat{\mathbb{C\,S}}_z(p)\longrightarrow{\mathbb{C\,S}}_z(p)$, it follows that $\psi_{(z)}$ acts as a conformal diffeomorphism between $\hat{\Sigma}_z$ and ${\Sigma}_z$
as long as the exponential maps are diffeomorphisms from $\widehat{\mathbb{C\,S}}_z(p)$ and ${\mathbb{C\,S}}_z(p)$ onto their respective images $\widehat{\Sigma}_z$ and ${\Sigma}_z$. Later we shall see how this result can be extended, under suitable hypotheses, to the less regular case of Lipschitzian exponential map. Here, we restrict our attention to  
the stated regularity assumptions on the exponential maps $\widehat{\exp}_p$ and $\exp_p$. They imply that the sky sections $\hat{\Sigma}_z$ and  ${\Sigma}_z$ have the topology of a 2-sphere. Moreover, we can take advantage of the fact that $(\widehat{\Sigma}_z,\,\widehat{g}_{(z)})$ is a (rescaled) round sphere, thus 
 we can apply the Poincar\'e--Koebe uniformization theorem, to the effect that there is a positive scalar function $\Phi_{\widehat\Sigma\,\Sigma}\,\in\,C^\infty(\widehat{\Sigma}_z, \mathbb{R}_{>0})$ such that
\begin{equation}
\label{psiconfbisse}
\left(\psi_{(z)}^*g_{(z)}\right)_{ab}\,=\,\frac{\partial\psi_{(z)}^i(y)}{\partial y^a}\frac{\partial\psi_{(z)}^k(y)}{\partial y^b}\,(g_{(z)})_{ik}\,=\,\Phi^2_{\widehat\Sigma\,\Sigma}\,(\widehat{g}_{(z)})_{ab}\,.
\end{equation} 
The required conformal factor $\Phi_{\widehat\Sigma\,\Sigma}\,\in\,C^\infty(\widehat{\Sigma}_z, \mathbb{R}_{>0})$ is the solution, (unique up to the $\mathrm{PSL}(2, \mathbb{C})$ action on $(\widehat\Sigma_z,\,\widehat{g}_{(z)})$), of the elliptic partial differential equation  on $(\widehat\Sigma_z, \hat{g}_{(z)})$ defined by \cite{Berger}  
\begin{equation}
\label{lapconf}
-\,\Delta _{\widehat{g}_{(z)}}\ln({\Phi}_{\widehat{\Sigma}\Sigma}
^2)\,+\,R(\widehat{g}_{(z)})\,=\, R(\psi_{(z)}^*g_{(z)})\,{\Phi}_{\widehat{\Sigma}\Sigma}
^2\;,
\end{equation}
where $\Delta_{\widehat{g}_{(z)}}\,:=\,\widehat{g}_{(z)}^{ab}\nabla_a\nabla_b$ is the Laplace-Beltrami operator on $(\widehat\Sigma_z, \hat{g}_{(z)})$, and where 
we respectively denoted by $R(\widehat{g}_{(z)})$ and  $R(\psi_{(z)}^*g_{(z)})$  the scalar curvature of the metrics $\widehat{g}_{(z)}$ and $\psi_{(z)}^*g_{(z)}$. 
Notice that the scalar curvature $R(\hat{g}_{(z)})$ is associated with the metric (\ref{metrichatzeta}) evaluated for $\widehat{r}\,=\,\widehat{r}(L)$ and hence is given by the constant $R(\hat{g}_{(z)})\,=\, \left[\frac{a^2_0}{(1\,+\,z)^2}\,f^2\left(\widehat{r}\right)\right]^{-1}$. Similarly, $R(g_{(z)})$ is associated with the metric (\ref{etametric0}) evaluated for 
$r\,=\,r(z)$, and as such it depends on the area distance $D^2(r(z), \theta,\phi)$ and  the lensing distortion $\mathcal{L}_{ab}$. \\
\\
By tracing (\ref{psiconfbisse}) with respect to $\widehat{g}_{(z)}^{ab}$, we get $tr_{\widehat{g}_{(z)}(y)}\,\left(\psi_{(z)}^* g_{(z)}\right)\,=\,2\Phi^2_{\widehat\Sigma\,\Sigma}$, 
and we can wite 
\begin{equation}
\label{Fdensity}
{\Phi}_{\widehat{\Sigma}\Sigma}^2\,=\,\frac{1}{2}\,tr_{\widehat{g}_{(z)}(y)}\,\left(\psi_{(z)}^* g_{(z)}\right)\,=\,\frac{1}{2}\,
\widehat{g}_{(z)}^{ab}\,\frac{\partial\psi_{(z)}^i(y)}{\partial y^a}\frac{\partial\psi_{(z)}^k(y)}{\partial y^b}\,(g_{(z)})_{ik}\;.
\end{equation}
From (\ref{psiconfbisse}) we also get $\det \left(\psi_{(z)}^* g_{(z)}\right)\,=\, {\Phi}_{\widehat{\Sigma}\Sigma}^4\,\det (\widehat{g}_{(z)})$, hence we can equivalently express the conformal factor ${\Phi}_{\widehat{\Sigma}\Sigma}^2$ as the  Radon-Nikodym derivative of the Riemannian measure $d\mu_{\psi^*{g}_{(z)}}\,:=\,\psi_{(z)}^*d\mu$ of  the pulled back metric $\psi_{(z)}^*g_{(z)}$ on the sky section $\widehat{\Sigma}_z$,  with respect to the Riemannian measure  $d\mu_{\widehat{g}_{(z)}}$ of the round metric $\widehat{g}_{(z)}$ on $\widehat\Sigma_z$,  \emph{i.e.}, 
\begin{equation}
\label{RadonNphi}
{\Phi}_{\widehat{\Sigma}\Sigma}^2\,=\,\frac{d\mu_{\psi_{(z)}^*{g_{(z)}}}}{d\mu_{\widehat{g}_{(z)}}}\,=\,
\frac{\psi_{(z)}^*d\mu_{g_{(z)}}}{d\mu_{\widehat{g}_{(z)}}}\;.
\end{equation}
Directly from this latter relation and from $E\left[\widehat{g}_{(z)},\, \psi_{(z)},\,{g}_{(z)}\right]\,=\,E\left[\widehat{h},\, \zeta_{(z)},\,{h}\right]$ (see (\ref{chainrelation})), we get
\begin{equation}
\label{Phirelation}
E\left[\widehat{h},\, \zeta_{(z)},\,{h}\right]\,=\,\int_{\widehat{\Sigma}_z}\,{\Phi}_{\widehat{\Sigma}\Sigma}^2\,d\mu_{\hat{h}}\,,
\end{equation}
which expresses the harmonic map functional $E\left[\widehat{h},\, \zeta_{(z)},\,{h}\right]$ in terms of the conformal factor ${\Phi}_{\widehat{\Sigma}\Sigma}^2$. As the $\mathrm{PSL}(2, \mathbb{C})$-localized map $\zeta_{(z)}$ varies with the scale $L(z)$, relation (\ref{Phirelation}) shows that $E\left[\widehat{h},\, \zeta_{(z)},\,{h}\right]$ describes the $\zeta_{(z)}$-dependent total "energy" associated with the conformal stretching of $(\widehat{\mathbb{C\,S}}_z(p),\,\widehat{h})$ over $(\mathbb{C\,S}_z(p),\,h)$.

\subsection{A local expression for ${\Phi}_{\widehat{\Sigma}\Sigma}^2$}
It is worthwhile to provide a local expression for ${\Phi}_{\widehat{\Sigma}\Sigma}^2$ showing the explicit dependence from the celestial coordinates $(\theta, \phi)$, the area distances $\widehat{D}(\widehat{r}(L))$, \, $D(r(L), \theta, \phi)$, and the distortion tensor $\mathcal{L}$ (see (\ref{etametric0})). We proceed as follows. Let us consider one of the coordinate bin $\widehat{B}(y_{(I)}, \delta)$ (see (\ref{hatbin})) in the celestial sphere $\widehat{\mathbb{C\,S}}_z(p)$. For $y=(r(z),\widehat\theta,\widehat\phi)\in\,\widehat{B}(y_{(I)}, \delta)$ let $q:=\widehat{\exp}_p(y)$ the point in the sky section $\widehat\Sigma_z$ reached, at the scale $L(z)$, along the past-directed null geodesics associated with the observational direction $y=(\widehat\theta, \widehat\phi)$. From the expression (\ref{RadonNphi}) of the conformal factor ${\Phi}_{\widehat{\Sigma}\Sigma}^2$ in terms of the measure $\psi_{(z)}^*d\mu_{g_{(z)}}$ we get, by massaging pull-backs, 
\begin{eqnarray}
\label{RadonNphi2}
{\Phi}_{\widehat{\Sigma}\Sigma}^2\,d\mu_{\widehat{g}_{(z)}}(q)\,&=&\,
\psi_{(z)}^*d\mu_{g_{(z)}}(q)\\
&=&\left(\exp_p\,\circ\,\zeta_{(z)}\,\circ\,\widehat{\exp}_p^{\,-1}\right)^*d\mu_{g_{(z)}}\nonumber\\
&=& (\widehat{\exp}_p^{\,-1})^*(\zeta_{(z)}^*d\mu_h)\nonumber\\
&\Rightarrow&\,\widehat{\exp}_p^*\left(
{\Phi}_{\widehat{\Sigma}\Sigma}^2\,d\mu_{\widehat{g}_{(z)}}(q)\right)\,
=\,\zeta_{(z)}^*d\mu_h (y)\nonumber\\
{\Phi}_{\widehat{\Sigma}\Sigma}^2(y)\,d\mu_{\widehat{h}}(y)\,&=&\,\zeta_{(z)}^*d\mu_h (y)\nonumber\,.
\end{eqnarray}
Hence, on $(\widehat{\mathbb{C\,S}}(p),\,\widehat{h})$, we need to compute the Radon-Nicodym derivative 
\begin{equation}
{\Phi}_{\widehat{\Sigma}\Sigma}^2(y)\,=\,
\frac{\zeta_{(z)}^*d\mu_h}{d\mu_{\widehat{h}}}(y)\,. 
\end{equation}
If we take into account the characterization $\sqrt{\det(h(r(z),\theta,\phi))}\,=\,D^2(r(z), \theta, \phi)\,\sqrt{\det(\widetilde{h}(\mathbb{S}^2))}$ of the area distance $D^2(r(z), \theta, \phi)$ (see (\ref{Di})), we compute
\begin{equation}
\zeta_{(z)}^*d\mu_h(y)\,=\,\left|\mathrm{Jac}_y(\zeta_{(z)})\right|\,
D^2(y)\,d\mu_{\mathbb{S}^2}\,,
\end{equation}
where $|\mathrm{Jac}_y(\zeta_{(z)})|$ is the Jacobian determinant associated with the localized $\mathrm{PSL}(2, \mathbb{C})$ map $\zeta_{(z)}$, \,and where $D^2(y)$ is a shorthand notation for 
the area distance $D^2(\zeta_{(z)}(\widehat{r}(z), \widehat\theta, \widehat\phi))$ pulled back  at $y\in (\widehat{\mathbb{C\,S}}_z(p),\,\widehat{h})$ by the localized $\zeta_{(z)}$. Similarly, from (\ref{Farea}) we compute
$d\mu_{\widehat{h}}(y)\,=\,\frac{a^2_0}{(1\,+\,z_{L})^2}\,f^2
\left(\widehat{r}(L)\right)\,d\mu_{\mathbb{S}^2}\,$. Thus, we can write
\begin{equation}
\label{expliPhi}
{\Phi}_{\widehat{\Sigma}\Sigma}^2(\widehat{r}(z),\widehat\theta,\widehat\phi)\,=\,
\left|\mathrm{Jac}\left(\zeta_{(z)}(\widehat{r}(z),\widehat\theta,\widehat\phi)\right)\right|\,
\frac{D^2(\zeta_{(z)}\left(\widehat{r}(z), \widehat\theta, \widehat\phi)\right)(1\,+\,z)^2}{a^2_0\,f^2
\left(\widehat{r}(z)\right)}\,.
\end{equation}
In terms of the FLRW area distance
\begin{equation}
\label{setLFLRWbis}
\widehat{D}(\widehat{r}(z))\,=\,
\frac{a_0}{1\,+\,z}\,f\left(\widehat{r}\right)\,,
\end{equation} 
we can equivalently write (\ref{expliPhi}) in the simpler form (where, to have handy the formula for later use, we have taken the square root)
\begin{equation}
\label{expliPhi2}
{\Phi}_{\widehat{\Sigma}\Sigma}(\widehat{r}(z),\widehat\theta,\widehat\phi)\,=\,
\left|\mathrm{Jac}\left(\zeta_{(z)}(\widehat{r}(z),\widehat\theta,\widehat\phi)\right)\right|^{\frac{1}{2}}\,
\frac{D\left(\zeta_{(z)}(\widehat{r}(z),\widehat\theta,\widehat\phi)\right)}{\widehat{D}(\widehat{r}(z))}\,.
\end{equation}
This clearly shows that the conformal factor ${\Phi}_{\widehat{\Sigma}\Sigma}$ is an explicit and, at least in principle, measurable quantity associated with the local Lorentz mapping (described by the localized  $\mathrm{PSL}(2,\mathbb{C})$ map $\zeta_{(z)}$) needed for adjusting the three reference null directions in the chosen celestial coordinates bin 
$\widehat{B}(y_{(I)}, \delta)$ in the celestial sphere $\widehat{\mathbb{C\,S}}_z(p)$. This adjustment allows to transfer to $\widehat{B}(y_{(I)}, \delta)$ the actual area distance, namely, compute $D(\zeta_{(z)}(\widehat{r}(z), \widehat\theta, \widehat\phi))$, and compare its  distribution on the FLRW celestial sphere $\widehat{\mathbb{C\,S}}_z(p)$ with respect to the isotropic   FLRW area distance $\widehat{D}(\widehat{r}(z))$. The  anisotropies in the angular distribution with respect to $\widehat{D}(\widehat{r}(z))$ give rise to fluctuations in ${\Phi}_{\widehat{\Sigma}\Sigma}$. It may appear somewhat surprising that, after all, the conformal factor does not explicitly depend also from the distortion tensor $\mathcal{L}_{ab}$ defined by (\ref{etametric0}). This dependence is implicit in the definition of the area distance (\ref{Di}) and of the coordinate parametrization  (\ref{etametric0}) characterizing $\mathcal{L}_{ab}$. These definitions give rise to the relation  (\ref{notracefree}) that, as can be easily checked, remove the explicit $\mathcal{L}_{ab}$ dependence from ${\Phi}_{\widehat{\Sigma}\Sigma}$. As we shall see, this fact will turn to our advantage when extending our analysis to the more general case of fractal-like sky sections.

\section{The sky section comparison functional at scale $L$}
The harmonic energy $E\left[\widehat{h},\, \zeta_{(z)},\,{h}\right]$, or equivalently 
$E\left[\widehat{g}_{(z)},\, \psi_{(z)},\,{g}_{(z)}\right]$, associated with the maps $\zeta_{(z)}$ and $\psi_{(z)}$, cannot be used directly as comparison functional between the sky sections $(\widehat\Sigma_z,\,\widehat{g}_{(z)})$ and $(\Sigma_z,\,{g}_{(z)})$. This follows directly as a consequence of the conformal invariance (\ref{conftrsf}) which implies\\

\begin{eqnarray}
&&E\left[\widehat{g}_{(z)},\, \psi_{(z)},\,{g}_{(z)}\right]\,=\,
\frac{1}{2}\,\int_{\widehat\Sigma_z}\,
(\widehat{g}_{(z)})^{ab}\,\frac{\partial\psi_{(z)}^i(y)}{\partial y^a}\frac{\partial\psi_{(z)}^k(y)}{\partial y^b}\,(g_{(z)})_{ik}\,d\mu_{\widehat{g}_{(z)}}
\\
&&=
\frac{1}{2}\,\int_{\widehat\Sigma_z}\,\left[\frac{a^2_0\,f^2
\left(\widehat{r}_{(z)}\right)}{(1\,+\,z_L)^2}\right]^{-1}
(\widehat{\widetilde{h}}(\mathbb{S}^2))^{ab}\,\frac{\partial\psi_{(z)}^i(y)}{\partial y^a}\frac{\partial\psi_{(z)}^k(y)}{\partial y^b}\,(g_{(z)})_{ik}\,
\left[\frac{a^2_0\,f^2
\left(\widehat{r}_{(z)}\right)}{(1\,+\,z_L)^2}\right]\,d\mu_{\mathbb{S}^2}\nonumber\\
&&=
\frac{1}{2}\,\int_{\widehat\Sigma_z}\,
(\widehat{\widetilde{h}}(\mathbb{S}^2))^{ab}\,\frac{\partial\psi_{(z)}^i(y)}{\partial y^a}\frac{\partial\psi_{(z)}^k(y)}{\partial y^b}\,(g_{(z)})_{ik}\,
d\mu_{\mathbb{S}^2}\nonumber\,,
\end{eqnarray}  
where, as usual, $\widehat{\widetilde{h}}(\mathbb{S}^2))$ is the round metric on the unit 2-sphere $\mathbb{S}^2$. From the above relation it follows that $E\left[\widehat{g}_{(z)},\, \psi_{(z)},\,{g}_{(z)}\right]$, (and similarly for $E\left[\widehat{h},\, \zeta_{(z)},\,{h}\right]$), does not depend from the area distance $\frac{a^2_0}{(1\,+\,z)^2}\,f^2
\left(\widehat{r}(z)\right)$ on the FLRW past lightcone $\widehat{\mathcal{C}}^-(p)$. Thus,  $E\left[\widehat{g}_{(z)},\, \psi_{(z)},\,{g}_{(z)}\right]$ cannot be a good candidate for the role of the functional that compares the sky sections $(\widehat\Sigma_z,\,\widehat{g}_{(z)})$ and $(\Sigma_z,\,{g}_{(z)})$. For this role, we introduced in \cite{CarFam} a functional whose structure was suggested by the rich repertoire of functionals used 
 in the problem of comparing shapes of surfaces in relation to computer graphic and visualization problems (see \textit{e.g.} \cite{JinYauGu} and \cite{YauGu}, to quote two  relevant papers in a vast literature). In particular, we were inspired by an energy functional introduced, under the name of \textit{elastic energy}, in a remarkable paper by J. Hass and P. Koehl  \cite{HassKoehl}, who use it as a powerful means  of comparing the shapes of genus-zero surfaces in problems relevant to surface visualization.\\
\\
In the more complex framework addressed in cosmography, we found it useful to define the sky section comparison functional at scale $L(z)$ according to
\begin{equation}
\label{enfunct1}
E_{\widehat\Sigma\Sigma}[\psi_{(z)}]\,:=\,\int_{\widehat{\Sigma}_z}({\Phi}_{\widehat\Sigma\Sigma}\,-\,1 )^2\,d\mu_{\hat{g}_{(z)}}\;,
\end{equation} 
that can be, more expressively, rewritten as  (see (\ref{expliPhi2}))
\begin{equation}
\label{variance}
E_{\widehat\Sigma\Sigma}[\psi_{(z)}]\,:=\,\int_{\widehat{\Sigma}_z}\,
\left[
\frac{\left|\mathrm{Jac}\left(\zeta_{(z)}(\widehat{r}(z))\right)\right|^{\frac{1}{2}}\,
D\left(\zeta_{(z)}(\widehat{r}(z),\widehat\theta,\widehat\phi)\right)\,-\,\widehat{D}(\widehat{r}(z))}{\widehat{D}(\widehat{r}(z))}\right]^2
\,d\mu_{\hat{g}_{(z)}}\;.
\end{equation}
Thus, from the physical point of view, $E_{\widehat\Sigma\Sigma}[\psi_{(z)}]$ describes the mean square fluctuations of the physical area distance  $D\left(\zeta_{(z)}(\widehat{r}(z),\widehat\theta,\widehat\phi)\right)$ (biased by the localized $\mathrm{PSL}(2, \mathbb{C})$ mapping $\zeta_{(z)}$) with respect to the reference FLRW isotropic area distance $\widehat{D}(\widehat{r}(z))$.\\
Notice that,  whereas the harmonic map energy $E\left[\widehat{g}_{(z)},\, \psi_{(z)},\,{g}_{(z)}\right]$ is a conformal invariant quantity,  the functional $E_{\widehat\Sigma\Sigma}[\psi_{(z)}]$ is not conformally invariant. Under a conformal transformation $\hat{h}\,\longrightarrow\, e^{2f}\,\hat{h}$ we get
\begin{equation}
\int_{\widehat{\Sigma}_z}\left(e^{\,-\,f}{\Phi}_{\widehat\Sigma\Sigma}\,-\,1 \right)^2\,e^{2f}\,d\mu_{\hat{h}}\;.
\end{equation}
Since we can also write
\begin{equation}
{\Phi}_{\widehat\Sigma\Sigma}\,=\,\left[\frac{\psi_{(z)}^*d\mu_{g_{(z)}}}{d\mu_{\widehat{g}_{(z)}}}   \right]^{\frac{1}{2}}\,,
\end{equation}
(see (\ref{RadonNphi})), it is also clear from its definition that corresponding to large linear "stretches" in conformally mapping $\psi_{(z)}^*g_{(z)}$ on $\widehat{g}_{(z)}$,  $E_{\widehat\Sigma\Sigma}[\psi_{(z)}]$ tends to the harmonic map energy.\\
\\
In our particular framework, the functional $E_{\widehat\Sigma\Sigma}[\psi_{(z)}]$ has many important properties that make  it a natural candidate for comparing, at the given length scale $L$, the sky sections $(\widehat\Sigma_z,\,\widehat{g}_{(z)})$ and $(\Sigma_z,\,{g}_{(z)})$ and, as the length-scale $L$ varies, the
physical lightcone region $\mathcal{C}^-_{L}(p,{g})$  with the FLRW reference region $\mathcal{C}^-_{L}(p,\hat{g})$.  These properties are discussed in detail in \cite{CarFam} (see Lemma 8 and Theorem 9), here we recollect them, without presenting their proof, in the following\footnote{In \cite{CarFam}, the general notation is somehow at variance from the one adopted here, since we address the analysis of $E_{\widehat\Sigma\Sigma}$ directly on the surfaces $\widehat\Sigma$ and $\Sigma$. In particular,  we refer to $\widehat\Sigma$ and $\Sigma$ as celestial spheres rather than sky sections.} 
\begin{theorem}
\label{propfunct}
The functional  $E_{\widehat\Sigma\Sigma}[\psi_{(z)}]$ is symmetric
\begin{equation}
\label{symmetryE}
E_{\widehat\Sigma\Sigma}[\psi_{(z)}]\,=\,E_{\Sigma\widehat\Sigma}[\psi^{-1}_{(z)}]\;,
\end{equation}
where
\begin{equation}
\label{invenfunct1}
E_{\Sigma\widehat\Sigma}[\psi^{-1}_{(z)}]\,:=\,\int_{{\Sigma}_z}({\Phi}_{\Sigma\widehat\Sigma}\,-\,1 )^2\,d\mu_{{g}_{(z)}}\;,
\end{equation}
 is the      comparison    functional associated with the inverse map $\psi_{(z)}^{-1}\,:\,{\Sigma}_z\,\longrightarrow\,\widehat\Sigma_z$, and ${\Phi}_{\Sigma\widehat\Sigma}$ is the corresponding conformal factor.\\
Let $(M,\,\widetilde{g})$ be another member of the FLRW family of spacetimes, distinct from $(M,\, \hat{g})$, that we may wish to use as a control in a best-fitting procedure for the physical spacetime $(M, g)$. Let $(\widetilde{\Sigma}_z,\, \widetilde{g}_{(z)})$ denote the sky section on the past 
lightcone  $\widetilde{\mathcal{C}}^-_{L_0}(p,\tilde{g})$, with vertex at $p$, and let  $\widetilde\psi_{(z)}\,:\Sigma_z\,\longmapsto\,\widetilde\Sigma_z$, and \,$\Phi_{\Sigma\widetilde\Sigma}$ respectively denote the corresponding diffeomorphism and conformal factor. Then to the composition of maps
\begin{equation}
\widehat{\Sigma}_z\,\underset{\psi_{(z)}}\longrightarrow\,\Sigma_z\,\underset{\widetilde\psi_{(z)}}\longrightarrow\,\widetilde\Sigma_z
\end{equation}
we can associate the triangular inequality
\begin{equation}
\label{triangularE}
E_{\widehat\Sigma\Sigma}[\psi_{(z)}]\,+\,E_{\Sigma\widetilde\Sigma}[\widetilde\psi_{(z)}]\,\geq\,
E_{\widehat\Sigma\widetilde\Sigma}[(\widetilde\psi_{(z)}\circ \psi_{(z)})]\,,
\end{equation}
where
 \begin{equation}
\label{invenfunct2tilde}
E_{\widehat\Sigma\widetilde\Sigma}[(\widetilde\psi_{(z)}\circ \psi_{(z)})]\,:=\,\int_{\widehat{\Sigma}_z}({\Phi}_{\widehat\Sigma\widetilde\Sigma}\,-\,1 )^2\,d\mu_{\widehat{g}_{(z)}}\;.
\end{equation}
Moreover,
\begin{equation}
E_{\widehat\Sigma\Sigma}[\psi_{(z)}]\,=\,0
\end{equation}
iff  the sky sections $(\widehat\Sigma,\,\hat{g}_{(z)})$ and  $(\Sigma,\,{g}_{(z)})$  are isometric.
Finally, if we denote by $\mathrm{W}^{1,2}_{\zeta_{(z)}} (\widehat{\mathbb{C\,S}}_z(p),\,\mathbb{C\,S}_z(p))$ the space of localized $\mathrm{PSL}(2,\mathbb{C})$- maps $\zeta_{(z)}$ which are of Sobolev class $\mathrm{W}^{1,2}$, (\emph{i.e.} square summable together with their first derivatives), then
\begin{equation}
d_{(z)}\left[\widehat{\Sigma}_z,\,\Sigma_z\right]\,:=\,\inf_{\zeta_{(z)}\in \mathrm{W}^{1,2}_{\zeta_{(z)}} (\widehat{\mathbb{C\,S}}_z(p),\,\mathbb{C\,S}_z(p))}\,E_{\widehat\Sigma\Sigma}[\psi_{(z)}]
\end{equation}
defines a scale-dependent   distance between the sky sections  $(\widehat{\Sigma}_z,\,\hat{g}_{(z)})$ and $(\Sigma_z,\, g_{(z)}) $ on the lightcone regions 
$\mathcal{C}^-_{L}(p,\hat{g})$ and $\mathcal{C}^-_{L}(p,g)$.
 \end{theorem}

We need to conclude our long lightcone journey addressing the real nature of the physical sky section $\Sigma_z$. This forces us to leave the comfort zone of the assumed smoothness of the past physical lightcone  ${\mathcal{C}}^-(p, \widehat{g})$.
\vskip 1cm
\section{The Lipschitz geometry of the cosmological sky sections $\Sigma_z$}
The celestial sphere description of the sky sections $\Sigma_z$ discussed above is inherently vulnerable to the vagaries of the local distribution of astrophysical sources, and the associated strong gravitational lensing phenomena\footnote{See \cite{Perlick} for a thorough analysis of the geometry of gravitational lensing.} imply that the actual past light cone $\mathscr{C}^-(p,g)$ is not smooth as we have assumed\footnote{The restrictive nature of the smoothness assumption on the metric $g$, typically represented by functions $g_{ab}\in C^k(\mathbb{R}^4, \mathbb{R})$,\,$k\geq 2$, and of the associated light cone, has been pointed out by many authors, mainly in the context of the proof of singularity theorems and in causality theory, (see \emph{e.g.} \cite{LeFloch}, \cite{Chrusciel},  \cite{Kunzinger},\cite{Minguzzi}, \cite{Senovilla}).}. In particular,  $\mathscr{C}^-(p,g)$ may fail to be the boundary $\partial\,\mathrm{I}^-(p,g)$ of the chronological past $\mathrm{I}^-(p,g)$ of $p$, (the set of all events $q\in M$ that can be connected to $p$ by a past-directed timelike curve), because past-directed null geodesics generators of $\mathscr{C}^-(p,g)$,\;  $\lambda\,:\,[0, \delta)\,\longrightarrow\,(M, g)$, 
with $\lambda(0)\,=\,p$,  may leave $\partial\mathrm{I}^-(p,g)$ and, under the action of the local spacetime curvature, plunge into the interior
 $\mathrm{I}^-(p,g)$. A spacetime description of this behaviour in connection with the phenomenology of gravitational lensing is discussed in detail in \cite{Perlick}, with a rich repertoire of examples of the possible singular structure that $\mathscr{C}^-(p,g)$ may  induce on the cosmological sky sections $\Sigma(p, r)$. As a matter of fact,  the sections $\Sigma_z$ may evolve into fractal-like surfaces, and to describe them from the point of view of geometric analysis, we need to introduce a framework tailored to the low-regularity landscape generated by the local inhomogeneities. \\

\subsection{The Lipschitz landscape} 
Given a past-directed null geodesic $I_W\,\ni\,r\longmapsto\exp_p(r k(n(\theta, \phi)))$, issued from $p\in\,M$ in the direction $n(\theta, \phi)\,\in\,\mathbb{C\,S}_z$, we follow \cite{Klainer} and define its \emph{terminal point} as the last\textendash{}point
\begin{equation}
q(r_*, n(\theta, \phi)):=\exp_p(r k(n(\theta, \phi)))
\end{equation}
that lies on the boundary $\partial\mathrm{I}^-(p,g)$ of the chronological past of $p$. Any such terminal point
$q(r_*, n(\theta, \phi))$ is said to be:\, \emph{i)}\; a \emph{conjugate terminal point} if the exponential map $\exp_p$ is singular at $\left(r_*,\,n(\theta,\phi)\right)$;\;\emph{ii)}\; a \emph{cut locus terminal point} 
if the exponential map $\exp_p$ is non--singular at $\left(r_*,\,n(\theta,\phi)\right)$ and there exists another null geodesic, issued from $p$, passing through $q(r_*, n(\theta, \phi))$, (see also \cite{Beem}, \cite{Perlick}). We denote \cite{Klainer} by $\mathcal{T}^-(p)$ the set of all terminal points associated with the past null geodesic flow issuing from $p$. In presence of cut points,  $\mathscr{C}^-(p,g)$ fails to be an embedded submanifold of $(M, g)$. Failure to be an immersed manifold is more directly related to conjugate points along the generators of $\mathscr{C}^-(p,g)$ and of the associated conjugate locus \cite{Perlick}. It follows that in presence of terminal points the mapping
\begin{equation}
\left.\exp_p\right|_{\mathscr{C}^-(p,g)}\,:\,\mathbb{C\,S}_z\,\longrightarrow\,
\Sigma_z\,:=\,\exp_p\,[\mathbb{C\,S}_z]
\end{equation} 
is no longer one-to-one, and the cosmological sky section $\Sigma_z$ fails to be a smooth surface. From the physical point of view, this is the geometrical setting associated with the generation of multiple  images of astrophysical sources\footnote{If the sources are not pointlike, we also have the more complex ring patterns typical of strong gravitational lensing.} in the observer celestial sphere  $\mathbb{C\,S}_z$. The mathematical framework for handling such a scenario is to assume that
the past null cone $\mathscr{C}^-(p,g)$ has the regularity of a Lipschitz manifold, characterized by a maximal atlas $\mathcal{A}\,=\,\left\{(U_\alpha, \varphi_\alpha)\right\}$ such that all transition maps between the coordinate charts $(U_\alpha, \varphi_\alpha)$ of $\mathscr{C}^-(p,g)$,
\begin{equation}
\varphi_{\alpha\beta}:=\varphi_\beta\circ\varphi_\alpha^{-1}\,:\,\varphi_\alpha\left(U_\alpha\cap\,U_\beta\right)\,
\longrightarrow\,\varphi_\beta\left(U_\alpha\cap\,U_\beta\right),\,
\end{equation}
are locally Lipschitz maps between domains of the Euclidean space $(\mathbb{R}^3, \delta)$. 
On $\mathscr{C}^-(p,g)$, the condition of being Lipschitz can be viewed as a weakened version of the differentiability. In particular, if $f:\,\mathscr{C}^-(p,g)\,\ni\,U\longrightarrow\mathbb{R}^3$ is a continuous map between open sets, then $f$ is Lipschitz if and only if it admits distributional partial derivatives that are in $L^\infty(U)$ with respect to the Lebesgue measure. This statement of Rademacher's theorem \cite{Gariepy}, \cite{Rosenberg} implies that the transition maps
$\varphi_{\alpha\beta}$ on $\mathscr{C}^-(p,g)$ have differentials $d\varphi_{\alpha\beta}$ that are defined almost everywhere, and which are locally bounded an measurable on their domains. In such a low-regularity setting the exponential map is quite delicate to handle. However, a key result, geometrically proved by M. Kunzinger, R. Steinbauer, M. Stojkovic \cite{Kuzinger}, (based on work by B.-L. Chen and P. LeFloch \cite{LeFloch}),  and by E. Minguzzi \cite{Minguzzi}, implies that the exponential map associated with a  $C^{1,\,1}$ metric can still be defined as a local bi-Lipschitz homeomorphism, namely a bijective map which along with its inverse is Lipschitz continuous in a sufficiently small neighborhood of $p$. Thus, the exponential map retains an appropriate form of regularity in the sense that locally, for each point $p\in\,M$, there exist open star-shaped neighborhoods, $N_0(p)$ of $0\in\,T_pM$ and $U_p\,\subset\,(M, g)$, such that $\exp_p\,:\,N_0(p)\,\longrightarrow\,U_p$ is a bi-Lipschitz homeomorphism \cite{Kuzinger}. In particular, each point $p\in (M, g)$ possesses a basis of totally normal neighborhoods. It is worthwhile to stress that geodesic normal coordinates (see (\ref{geodnormal})) can be still defined, but the transition from the current smooth coordinate systems\footnote{Recall that $M$ is a smooth manifold, and that the low Lipschitz $C^{1,\,1}$ regularity is caused by the metric $g$, and not by the differentiable structure of $M$.} used around $p\in M$ to the normal coordinates associated with $\exp_p$ is only continuous.\\

\subsection{The fractal-like sky section $\Sigma_z$}
We are interested in the geometry that such past light cone scenario induces on the cosmological sky section $\Sigma_z\,:=\,\exp_p\,[\mathbb{C\,S}_z]$ of $\mathcal{C}^-(p,g)$. 
As long as $\exp_p$ is bi-Lipschitz, the sky sections $\Sigma_z$ are topological 2-spheres, and the results above seem to suggest that after all there is no such a strong motivation to abandon the comforts of the smooth framework in favor of a Lipschitzian rugged landscape. However,  as the length scale $L$ varies,  the development of caustics in $\mathscr{C}^-(p,g)$ generates cusps and crossings in the surfaces $\Sigma_z$, to the effect that they  are no longer homeomorphic to 2-spheres. In such a setting, 
  the restriction of the exponential map to the celestial sphere $\mathbb{C\,S}_z$,  characterizing the surface $\Sigma_z$, (see (\ref{sigmapr})),
\begin{equation}
\label{bilipexp}
\exp_p\,:\,\mathbb{C\,S}_z\,\subset\,T_pM\,\longrightarrow\,\Sigma_z\,:=\,
\exp_p\left[\mathbb{C\,S}_z\right]\,\subset\,\mathscr{C}^-(p,g)\,,
\end{equation} 
is only a Lipschitz map between the metric spaces $\left(\mathbb{C\,S}_z,\,d_{\mathbb{S}_r^2} \right)$ and $\left(\Sigma_z,\,d_{g|\Sigma}\right)$, where $d_{\mathbb{S}_r^2}$ is the standard distance function on the round 2-sphere ${\mathbb{S}^2_r}$ or radius $r$, and $d_{g|\Sigma}$ is the distance function induced (almost everywhere) on $\Sigma_z$ by the metric $g|_{\Sigma_z}$ defined\footnote{In presence of cut points the inclusion map 
$\iota_r:\Sigma_z\,\hookrightarrow\,\mathscr{C}^-(p, g)$ of the sky section $\Sigma_z$  into $\mathscr{C}^-(p, g)$ is Lipschitz, thus Rademacher's theorem allows us to define the pull-back metric $g|_{\Sigma_z}\,:=\,\iota_r^*\,\left.g\right|_{\mathscr{C}^-(p, g)}$ only almost-everywhere} by  (\ref{Sigmametric}). In general, the sky section $\Sigma_z$ can be topologically very complex since it may contain terminal points of the exponential map $\exp_p$, giving rise to cusps and  swallow-tail points associated with self-intersections of $\Sigma_z$. Even if this may evolve in a very complex picture of $\Sigma_z$, we still have quite a geometric control over its metric structure. The Lipschitz regularity of $\exp_p$ implies that 
there is a constant $c_r$, depending on the parameter $r$, such that
\begin{equation}
\label{lipexp2}
d_{\Sigma(p, r)}\left(\exp_p(x), \exp_p(y)\right)\,\leq\,c_r\,d_{\mathbb{S}^2_r}(x, y),\,\,\,\,\forall\,x,\,y\,\in\,\mathbb{S}^2_r\,,
\end{equation}  
and we can define the pull-back on the celestial sphere $\mathbb{C\,S}_z\,\in\,T_pM$ of the distance function $d_{\Sigma_z}$ according to 
\begin{equation}
\exp_p^*d_{\Sigma_z}\,=\,
d_{g\Sigma}\left(\exp_p(x), \exp_p(y)\right)\,,\,\,\,\,\forall\,x,\,y\,\in\,\mathbb{C\,S}_z\,.
\end{equation}
We can also pull-back the metric $g|_{g|\Sigma_z}$ to $\mathbb{C\,S}_z$. 
By Rademacher's theorem $\exp_p$ is differentiable almost everywhere, and 
\begin{equation}
\label{metrichLip1}
h(\theta, \phi)\,:=\,\left(\exp_p^*\,g|_{\Sigma_z}\right)_{\alpha\beta}\,dx^\alpha dx^\beta\,,
\end{equation}
is a metric defined, almost everywhere on the celestial sphere $\mathbb{C\,S}_z$, 
(by a slight abuse of language, we have used the same notation as for the smooth version(\ref{metrich})). We can also define almost everywhere the volume element $d\mu_{h}$ associated with the metric (\ref{metrichLip1}), \emph{i.e.}
\begin{equation}
\label{hvolumeLip}
d\mu_{h}\,:=\,\exp_p^*d\mu_{g|_{\Sigma_z}}\,=\,\sqrt{\det(h(r(z),\theta,\phi))}\,d\theta d\varphi\,,
\end{equation} 
in full analogy with its smooth version (\ref{hvolume}). All this implies that with the proviso of the almost everywhere meaning, the
characterization (\ref{Di}) of the angular diameter distance $D(r, \theta, \phi)$ and of the shear-inducing distortion $L_{\alpha\beta}$ defined by (\ref{etametric0}), carry over to the bi-Lipschitz case.\\
\\
To put these geometrical remarks at work, let us stress that we cannot have reasonable control over the very complex topological structure of the sky section $\Sigma_z$ induced by a cascade of (strong) lensing events. Moreover, the corresponding caustics and singularities at the terminal points on $\Sigma_z$ provide a level of detail that is not relevant to the present analysis. Thus, as a reasonable compromise, we assume that the exponential map $\exp_p$ is bi-Lipschitz, that $\Sigma_z$ is topologically a 2-sphere, and we mimic the effect of the many lensing events that may affect $\Sigma_z$ by assuming that the sky section 
$\Sigma_z$ has the irregularities of a metric surface with the fractal geometry of a 2-sphere with the locally-finite Hausdorff 2-measure associated with (\ref{hvolumeLip}). Under such assumptions, it can be shown that our smooth analysis can be safely extended, (in particular, we can still exploit the Poincar\'e--Koebe uniformization theorem \cite{Dimitrios}), and the results obtained 
hold also in the more general setting of a Lipschitz description of the cosmographic past lightcone $\mathcal{C}^-(p,g)$.

\section{Concluding remarks: $d_{(z)}\left[\widehat{\Sigma}_z,\,\Sigma_z\right]$  as a scale-dependent field}
According to the physical characterization (\ref{variance}) of $E_{\widehat\Sigma\Sigma}[\psi_{(z)}]$, and the results described in Theorem \ref{propfunct},   the distance function $d_{(z)}\left[\widehat{\Sigma}_z,\,\Sigma_z\right]$, (for simplicity, 
one may work with the $E_{\widehat\Sigma\Sigma}[\psi_{(z)}]$ realizing the minimum), can be interpreted as defining a  $z$-dependent field on the FLRW past light cone $\widehat{\mathcal{C}}^-(p, \widehat{g})$ describing the mean square fluctuations of the anisotropies of the physical area distance $D(\zeta_{(z)})$ with respect to the reference FLRW area distance $\widehat{D}(\widehat{r}(z))$. These fluctuations provide information on how much the local area element on the physical sky section $\Sigma_z$ differs from the corresponding (round) area element on the reference FLRW sky section $\Sigma_z$. Since  for 2-dimensional surfaces the local Riemannian geometry is fully described by the area element, the fluctuations in $D(\zeta_{(z)})$ give information on how much the geometries of the sky sections $\widehat\Sigma_z$ and $\Sigma_z$ differ. When we reach the scale of homogeneity, the physical area distance
$D(\zeta_{(z)})$ becomes isotropic and can be identified with the reference FLRW $\widehat{D}(\widehat{r}(z))$. The localized null-directions alignment between the corresponding celestial spheres $\mathbb{C\,S}_z(p)$ and  $\widehat{\mathbb{C\,S}}_z(p)$ reduces to a global kinematical Lorentz boost (and a rotation). Thus, corresponding to this homogeneity scale, the distance function $d_{(z)}\left[\widehat{\Sigma}_z,\,\Sigma_z\right]$ field  vanishes.\\
\\
Thus, we have an interesting scenario whereby it is possible to associate with the distance functional $d_{(z)}\left[\widehat{\Sigma}_z,\,\Sigma_z\right]$ a scale-dependent field that describes a global effect that the reference FLRW  past lightcone $\widehat{\mathcal{C}}^-(p, \widehat{g})$ misses in describing the pre-homogeneity anisotropies of the actual past lightcone ${\mathcal{C}}^-(p, \widehat{g})$. This \emph{pre-homogeneity field} is, in line of principle, measurable since it is the mean-square variation of the physical area distance $D(\zeta_{(z)})$. The delicate question concerns its possible role in selecting the large-scale FLRW model that best fits the cosmological observations on large scales. A few qualitative indications in this direction, mainly of a perturbative nature, are discussed in \cite{CarFam}. The results presented here are however more precise since they connect directly the distance functional $d_{(z)}\left[\widehat{\Sigma}_z,\,\Sigma_z\right]$ to the area distance $D(\zeta_{(z)})$. To describe an important consequence of these results, 
let us consider the light cone regions  $\mathcal{C}^-_{L}(p,\hat{g})$ and $\mathcal{C}^-_{L}(p,g)$ over a sufficiently small length scale $L(z)$.
 If $\zeta_{(z)}$ and the corresponding $\psi_{(z)}$ denote the minimizing maps  characterized in Theorem \ref{propfunct}, then we can write \cite{CarFam}
\begin{eqnarray}
\label{enfunct2}
E_{\widehat\Sigma\Sigma}[\psi_{(z)}]\,&=&\,\int_{\widehat{\Sigma}_z}( {\Phi}_{\widehat\Sigma\Sigma}\,-\,1 )^2\,d\mu_{\widehat{g}_{(z)}}\,=\,
\int_{\widehat{\Sigma}_z}{\Phi}_{\widehat\Sigma\Sigma}^2\,d\mu_{\widehat{g}_{(z)}}\,+\,
\int_{\widehat{\Sigma}_z}\,d\mu_{\widehat{g}_{(z)}}\,-\,2\int_{\widehat{\Sigma}_z}{\Phi}_{\widehat\Sigma\Sigma}\,d\mu_{\widehat{g}_{(z)}}\nonumber\\
\\
&=&\int_{\widehat{\Sigma}_z}\frac{\psi_{g_{(z)}}^*d\mu_{g_{(z)}}}{d\mu_{\widehat{g}_{(z)}}}\,d\mu_{\widehat{g}_{(z)}}\,+\,
A\left(\widehat{\Sigma}_z\right)\,-\,2\int_{\widehat{\Sigma}_z}{\Phi}_{\widehat\Sigma\Sigma}\,d\mu_{\widehat{g}_{(z)}}\nonumber\\
&=&\int_{\psi_{(z)}(\widehat{\Sigma}_z)}{d\mu_{g_{(z)}}}\,+\,
A\left(\widehat{\Sigma}_z\right)\,-\,2\int_{\widehat{\Sigma}_z}{\Phi}_{\widehat\Sigma\Sigma}\,d\mu_{\widehat{g}_{(z)}}\nonumber\\
&=&A({\Sigma}_z)\,+\,
A\left(\widehat{\Sigma}_z\right)\,-\,2\int_{\widehat{\Sigma}_z}{\Phi}_{\widehat\Sigma\Sigma}\,d\mu_{\widehat{g}_{(z)}}\;,\nonumber
\end{eqnarray}
where we have exploited  the Radon-Nikodyn characterization of  $\widehat{\Phi}_{\widehat\Sigma\Sigma}^2$, (see (\ref{RadonNphi})), the identification $\psi_{(z)}(\widehat{\Sigma}_z)\,=\,\Sigma_z$,\, and the relation
\begin{equation}
\label{AreaSigma}
\int_{\widehat{\Sigma}_z}\frac{\psi_{(z)}^*d\mu_{{g}_{(z)}}}{d\mu_{\widehat{g}_{(z)}}}\,
d\mu_{\widehat{g}_{(z)}}=
\int_{\widehat{\Sigma}_z}{\psi_{(z)}^*d\mu_{g_{(z)}}}=\int_{\psi(\widehat{\Sigma}_z)}{d\mu_{g_{(z)}}}=\int_{{\Sigma}_z}{d\mu_{g_{(z)}}}\,=\,A({\Sigma}_z)\;,
\end{equation}
where $A({\Sigma}_z)$ and  $A\left(\widehat{\Sigma}_z\right)$ respectively denote the area of the sky sections  
$(\hat{\Sigma}_z,\,\widehat{g}_{(z)})$ and $(\Sigma_z,\, g_{(z)})$. Thus, 
 \begin{equation}
\label{intdist}
d_{(z)}\left[\widehat{\Sigma}_L,\,{\Sigma}_L  \right]\,=\,E_{\widehat\Sigma\Sigma}[\psi_{(z)}]\,:=\,A\left(\widehat{\Sigma}_z\right)\,+\,A({\Sigma}_z)\,
-\,2\int_{\widehat\Sigma_z}{\Phi}_{\widehat\Sigma\Sigma}\,d\mu_{\widehat{g}_{(z)}}\,.
\end{equation}
To simplify matters, we assume that at the given length scale $L(z)$ the corresponding region $\mathcal{C}^-_{L}(p,g)$ is caustic free. Let us rewrite   ${\Phi}_{\Sigma\widehat\Sigma}$ as
\begin{eqnarray}
{\Phi}_{\Sigma\widehat\Sigma}\,&=&\,\left({\Phi}_{\Sigma\widehat\Sigma}\,-\,1\right)\,+\,1
\\
&=&\,
\frac{\left|\mathrm{Jac}\left(\zeta_{(z)}\right)\right|^{\frac{1}{2}}\,D(\zeta_{(z)})\,-\,\widehat{D}(\widehat{r}(z))}{\widehat{D}(\widehat{r}(z))}\,+\,1\nonumber\,,
\end{eqnarray}
where we have simplified the notation used in (\ref{expliPhi2}). By introducing this in (\ref{intdist}) we get 
\begin{equation}
\label{distapprox0}
d_L\left[\widehat{\Sigma}_z,\,{\Sigma}_z  \right]\,=\,A\left({\Sigma}_z\right)\,-\,A(\widehat{\Sigma}_z)\,
-\,2\int_{\widehat\Sigma_z}\,\left[\frac{\left|\mathrm{Jac}\left(\zeta_{(z)}\right)\right|^{\frac{1}{2}}\,D(\zeta_{(z)})\,-\,\widehat{D}(\widehat{r}(z))}{\widehat{D}(\widehat{r}(z))}   \right]      \,d\mu_{\widehat{g}_{(z)}}\,.
\end{equation} 
This expression can be further specialized if  we exploit the asymptotic expressions of the area $A\left(\widehat{\Sigma}_z\right)$ and 
$A\left({\Sigma}_z\right)$ of the two surfaces $(\widehat{\Sigma}_z, \,\widehat{g}_{(z)})$, \,$({\Sigma}_z, \,{g}_{(z)})$ on the corresponding lightcones
$\mathcal{C}^-_{{L}}(p, \widehat{g})$ and  $\mathcal{C}^-_{{L}}(p, {g})$. These asymptotic expressions can be obtained if we consider the associated causal past regions $\mathcal{J}^-_{{L}}(p, \widehat{g})$  and $\mathcal{J}^-_{{L}}(p, {g})$ sufficiently near the (common) observation point $p$, in particular when the length scale $L(z)$ we are probing is small with respect to the "cosmological" curvature scale.  Under such assumption, there is a unique maximal 3-dimensional region $V_L^3(p)$, embedded in   $\mathcal{J}^-_{{L}}(p, {g})$, having the surface  $({\Sigma}_z, \,{h})$ as its boundary. This surface intersects the world line $\gamma(\tau)$ of the observer $p$ at the point $q=\gamma(\tau_0\,=\,-\,L(z))$ defined by the given length scale $L(z)$. For the reference FLRW the analogous set up is associated to the constant-time slicing of the FLRW spacetime $(M, \widehat{g})$ considered. The corresponding  3-dimensional region $\widehat{V}_L^3(p)$, embedded in   $\mathcal{J}^-_{{L}}(p, \widehat{g})$, has the surface  $(\widehat{\Sigma}_z, \,\hat{h})$ as its boundary.  The FLRW observer $\widehat{\gamma}(\widehat\tau)$ will intersect $\widehat{V}_L^3(p)$ at the point $\widehat{q}=\widehat{\gamma}(\widehat{\tau}_0\,=\,-\,L(z))$.  By introducing  geodesic normal coordinates $\{X^i\}$ in $\mathcal{J}^-_{{L}}(p, {g})$  and  $\{Y^k\}$ in $\mathcal{J}^-_{{L}}(p, \widehat{g})$, respectively based at the point $q$ and $\widehat{q}$, we can pull back the  metric tensors $g$ and $\widehat{g}$ to $T_{q}M$ and $T_{\widehat{q}}M$, and obtain the classical normal coordinate development of the metrics
$g$ and $\widehat{g}$ valid in a sufficiently small convex neighborhood of  $q$ and $\widehat{q}$. Explicitly, for the (more relevant case of the)  metric $g$, we have   (see \textit{e. g.} Lemma 3.4 (p. 210) of  \cite{SchoenYau} or \cite{Petersen}) 
\begin{eqnarray}
&&\left((\mathrm{exp}_q)^*\,g \right)_{ef}\,=\,\eta_{ef}\,-\,\frac{1}{3}\,\mathrm{R}_{eabf}|_qX^aX^b\,-\,\frac{1}{6}\,\nabla_c\mathrm{R}_{eabf}|_qX^aX^bX^c\nonumber\\
\nonumber\\
&&+\,\left(-\,\frac{1}{20}\,\nabla_c\nabla_d\mathrm{R}_{eabf}\,+\,\frac{2}{45}\,\mathrm{R}_{eabm}\,\mathrm{R}^m_{fcd} \right)_q\,X^aX^bX^cX^d\,+\,\ldots\nonumber\;,
\end{eqnarray}
where $\mathrm{R}_{abcd}$ is the Riemann tensor of the metric $g$ (evaluated at the point 
$q$).  The induced expansion in the pulled-back  measure $\left((\mathrm{exp}_{s(\eta)})^*d\mu_{g}\right)$ provides the Lorentzian analog of the familiar Bertrand-Puiseux formulas associated with   the geometrical interpretation of the sectional, Ricci and scalar curvature for a Riemannian manifold in terms of the length, area, and volume measures of small geodesic balls. In the Lorentzian case the relevant formulas are more delicate to derive, \cite{Berthiere}, \cite{GibbSol1}, \cite{GibbSol2}, \cite{Myrheim}. This asymptotics provides \cite{GibbSol1}, to leading order in $L(z)$, the following expressions for the area  of  $({\Sigma}_z, \,{g}_{(z)})$ and  
$(\widehat{\Sigma}_z, \,\widehat{g}_{(z)})$, 
\begin{equation} 
\label{Areaasymp1}
A\left({\Sigma}_z\right)\,=\,{\pi}\,L^2(z)\,\left(1\,-\,\frac{1}{72}\,L^2(z)\,\mathrm{R}(q)\,+\,\ldots  \right)\;,
\end{equation}
and
\begin{equation}
\label{Areaasymp1}
A\left(\widehat{\Sigma}_z\right)\,=\,{\pi}\,L^2(z)\,\left(1\,-\,\frac{1}{72}\,L^2(z)\,\widehat{\mathrm{R}}(\widehat{q})\,+\,\ldots  \right)\;,
\end{equation}
Introducing these expressions in (\ref{distapprox0}) we eventually get
\begin{equation}
\label{Rdapprox}
\widehat{\mathrm{R}}(\hat{q})\,=\,{\mathrm{R}}({q})\,+\,\frac{72}{\pi}\frac{d_{(z)}\left[\widehat{\Sigma}_z,\,{\Sigma}_z  \right]}{L^4(z)}\,
+\,\frac{144}{\pi L^4(z)}\,\int_{\widehat\Sigma_z}\,\left[\frac{\left|\mathrm{Jac}\left(\zeta_{(z)}\right)\right|^{\frac{1}{2}}\,D(\zeta_{(z)})\,-\,\widehat{D}(\widehat{r}(z))}{\widehat{D}(\widehat{r}(z))}   \right]\,d\mu_{\widehat{g}_{(z)}}\,+\,\ldots\,.
\end{equation}
Notice that the integral is the average value over the sky section $(\widehat\Sigma_z, \widehat{g}_{(z)})$, of the fluctuations of $\left|\mathrm{Jac}\left(\zeta_{(z)}\right)\right|^{\frac{1}{2}}\,D(\zeta_{(z)})$ with respect to $\widehat{D}(\widehat{r}(z))$, average that for notational ease we write as   
\begin{equation}
\left\langle\left.D(\zeta_{(z)})\right|\widehat{D}(\widehat{r}(z))\right\rangle_{\widehat\Sigma_z}\,:=\,A^{-1}(\widehat\Sigma_z)\,
\int_{\widehat\Sigma_z}\,\left[\frac{\left|\mathrm{Jac}\left(\zeta_{(z)}\right)\right|^{\frac{1}{2}}\,D(\zeta_{(z)})\,-\,\widehat{D}(\widehat{r}(z))}{\widehat{D}(\widehat{r}(z))}   \right]\,d\mu_{\widehat{g}_{(z)}}\,,
\end{equation}
while, as we have already stressed, the distance functional is (up to the  $A(\widehat\Sigma_z)$ normalization) the square mean deviation of this average, \emph{i. e.},
\begin{eqnarray}
\left\langle\left(\left.D(\zeta_{(z)})\right|\widehat{D}(\widehat{r}(z))\right)^2\right\rangle_{\widehat\Sigma_z}\,:\,&=&\,A^{-1}(\widehat\Sigma_z)\,
\int_{\widehat\Sigma_z}\,\left[\frac{\left|\mathrm{Jac}\left(\zeta_{(z)}\right)\right|^{\frac{1}{2}}\,D(\zeta_{(z)})\,-\,\widehat{D}(\widehat{r}(z))}{\widehat{D}(\widehat{r}(z))}   \right]^2\,d\mu_{\widehat{g}_{(z)}}\nonumber\\
\\
&=&\, A^{-1}(\widehat\Sigma_z)\,d_{(z)}\left[\widehat{\Sigma}_z,\,{\Sigma}_z  \right]\,.\nonumber
\end{eqnarray}
To put these results at work, let us assume the  conservative and quite a reasonable scenario where the fluctuations in the area distance $D(\zeta_{(z)})$, even if locally large in the various celestial coordinates bins, average out to zero over $\widehat{\Sigma}_z$. However, the corresponding square mean deviation of the fluctuations $\left\langle\left(\left.D(\zeta_{(z)})\right|\widehat{D}(\widehat{r}(z))\right)^2\right\rangle_{\widehat\Sigma_z}\,=\,A^{-1}(\widehat\Sigma_z)\,d_{(z)}\left[\widehat{\Sigma}_z,\,{\Sigma}_z  \right]$ can be significantly different from zero, and  from (\ref{Rdapprox})  we get
\begin{equation}
\label{Rdapprox2}
\widehat{\mathrm{R}}(\hat{q})\,=\,{\mathrm{R}}({q})\,+\,\frac{72}{\pi}\frac{d_{(z)}\left[\widehat{\Sigma}_z,\,{\Sigma}_z  \right]}{L^4(z)}\,+\,\ldots\,. 
\end{equation}
  The physical scalar curvature we measure (hard to!) in such a scenario is ${\mathrm{R}}({q})$, and if we decide to modeling with a FLRW solution a cosmological spacetime, homogeneous on large scale but highly inhomogeneous at smaller scale, then
(\ref{Rdapprox2}) shows that we cannot identify ${\mathrm{R}}({q})$ with the corresponding FLRW scalar curvature $\widehat{\mathrm{R}}(\hat{q})$. Such an identification is possible, with a rigorous level of scale dependence precision, only if we take into account the term 
\begin{equation}
\frac{72}{\pi}\frac{d_{(z)}\left[\widehat{\Sigma}_z,\,{\Sigma}_z  \right]}{L^4(z)}\,.
\end{equation}
According to Theorem \ref{propfunct}, this term vanishes once $L(z)$ probes the homogeneity scales, conversely, it is clear from (\ref{Rdapprox2}) that in pre-homogeneity region its presence is forced on us and  plays the role of a scale-dependent effective positive contribution to  the cosmological constant. As long as the local inhomogeneities give rise to significant fluctuations in the area distance $D(\zeta_{(z)})$, this contribution cannot be considered a priori negligible in high-precision cosmology.

\end{document}